\documentstyle[eqsecnum,epsfig,aps,preprint,tighten]{revtex}

\begin{document}
\draft
\preprint{IHEP 2000-39}
\title{Global fit to the charged leptons DIS data:
\boldmath{$\alpha_{\rm s}$,} parton distributions, and high twists}
\author{S. I. Alekhin}
\address{Institute for High Energy Physics, Protvino, 142284, Russia}
\date{October 2000}
\maketitle
\begin{abstract}
We perform the NLO QCD analysis of the world data 
on inclusive deep inelastic scattering cross sections 
of charged leptons off the proton and the deuterium targets. 
The parton distributions, the value of strong coupling constant
$\alpha_{\rm s}$, 
and the twist 4 contributions to the structure functions 
$F_2$ and $F_{\rm L}$ are extracted with the complete
account for the correlations of data points due to the systematic errors.
Sensitivity of the $\alpha_{\rm s}$ value and the high twist 
contribution to the procedures of accounting for 
the systematic errors is studied.
The impact of theoretical uncertainties 
on the value of $\alpha_{\rm s}$ and on the parton distributions is 
analysed. The obtained value of strong coupling constant
with the account of these uncertainties is 
$\alpha_{\rm s}(M_{\rm Z})=0.1165\pm0.0017({\rm stat+syst})
\pm^{0.0026}_{0.0034}({\rm theor})$. The uncertainties of 
parton-parton luminosities for the FNAL and LHC colliders 
are estimated.
\end{abstract}


\section{Introduction}
\label{introduction}

Experiments on deep inelastic scattering (DIS) of leptons off nucleons is 
a unique source of information about strong interaction.
These experiments were initiated at SLAC linac and later were continued
at different accelerators using the fixed targets and the colliding 
electron-proton beams. The data  
for proton and deuterium targets, given in 
Refs.~\cite{Whitlow:1992uw,Benvenuti:1989rh,Arneodo:1997qe,Adams:1996gu}
are especially valuable, since no heavy-nucleus corrections are needed
for their interpretation. Those data combined with the 
results from HERA electron-proton collider 
\cite{Aid:1996au,Derrick:1996hn} allows one to determine 
the parton distribution functions (PDFs) and 
are widely used for this purpose. In particular, global 
fits of PDFs, which are regularly updated by 
collaborations MRST \cite{Martin:2000ww}
and CTEQ\cite{Lai:2000wy}, rely heavily on the DIS data.
It is often mentioned, that the MRST and CTEQ PDFs lack information on 
uncertainties, that does not allow one to estimate the uncertainties 
on the cross sections, which are calculated using those PDFs.
Most often these uncertainties are estimated as a spread of results,
obtained using different PDFs sets. Meanwhile, it is evident, that  
if different PDFs are based on the same theoretical model
fitted to a similar data sets, this spread 
mainly reflects uncertainties of calculations, rather, than
real uncertainties arising from statistical and systematic 
errors on the data used for the extraction of PDFs.
Besides, those collaborations combine statistical and systematic errors
in quadrature, i.e. do not account for the correlation of the latter. 
Since systematic errors dominate over statistical ones for many DIS experiments, 
they govern total experimental errors on the 
PDFs parameters fitted to the data and ignorance of their correlations 
may lead to the distortion of the parameters errors and to the bias 
of their central values. 

Statistical and systematic errors are combined in quadrature 
in part by historical reasons. The other reason is that,
contrary to the case of statistical errors,
existing approaches to the account of systematic errors are not so 
straightforward and encounter with technical difficulties 
generated by correlations between measurements, which 
become more significant when the systematic errors rise,
as compared with the statistical ones. Nevertheless, as it was shown 
in Ref.\cite{Alekhin:1999za} on the example of combined analysis 
of DIS data from 
Refs.\cite{Benvenuti:1989rh,Arneodo:1997qe,Aid:1996au,Derrick:1996hn}
with the complete account of correlations due to systematic errors,
these difficulties can be overcomed using in the fit 
an estimator based on the covariance matrix. 
The results of the combined analysis of data from 
Refs.\cite{Whitlow:1992uw,Benvenuti:1989rh,Arneodo:1997qe,Adams:1996gu,Aid:1996au,Derrick:1996hn}, 
which attempted to account for correlations of systematic errors, was 
later given in Ref.\cite{Botje:2000dj},
but due to the large number of 
independent sources of the systematic errors, they were  
combined with the statistical errors partially. Regardless of the expressed confidence 
that this procedure should have minimal impact
on the results, this point is not ultimately clarified and 
it is evident that errors on the obtained PDFs may be distorted.

In this paper we describe the results of the combined analysis 
of the world data on the charged leptons DIS off the proton and deuterium 
targets given in Refs.  
\cite{Whitlow:1992uw,Benvenuti:1989rh,Arneodo:1997qe,Adams:1996gu,Aid:1996au,Derrick:1996hn}.
In comparison 
with our previous fit of Ref.\cite{Alekhin:1999za} in the present analysis we 
use data with lower values of transferred momentum $Q$.
Besides, the data from the SLAC experiments and the experiment FNAL-E-665 
are added. As well as in Ref.\cite{Alekhin:1999za}, we extract from the data 
the nucleon PDFs and the value of strong coupling constant 
$\alpha_{\rm s}$. In addition, wealth of data at low $Q$ allows one to determine 
the high twist (HT) contributions to the structure functions
$F_2$ and $F_{\rm L}$ as well.
Analysis is performed in the NLO QCD approximation with the complete 
account of correlations due to systematic errors within approach 
described in Ref.\cite{Alekhin:2000es}.

\section{The DIS phenomenology}
\label{sec:theory}

It is well known that the DIS cross section of charged leptons off 
nucleons can be expressed in terms of structure functions 
$F_{2,3,{\rm L}}$\footnote{The comprehensive analysis of 
lepton-nucleon scattering amplitudes, including notations used 
throughout our paper is given, e.g., in review \protect\cite{Altarelli:1982ax}.}.
For example, at 4-momentum transfers $Q$ less, than the  
$Z$-boson mass the charged leptons cross section reads
\begin{equation}
\frac{d^2\sigma}{dxdy}=\frac{4\pi\alpha^2(s-M^2)}{Q^4}
\left[\left(1-y-\frac{(Mxy)^2}{Q^2}\right)F_2+
\left(1-2\frac{m_{\rm l}^2}{Q^2}\right)
\frac{y^2}{2}\left(F_2-F_{\rm L}\right)\right],
\label{eqn:discs}
\end{equation}
where $s$ is the s.c.m. energy, $m_{\rm l}$ is the lepton mass,
$y$ is the ratio of the energy lost by lepton to
the initial lepton energy,
$x$ is the Bjorken scaling variable, 
$M$ is the nucleon mass,
$\alpha$ is the electro-magnetic coupling constant.
The structure functions 
$F_{2,\rm L}$ depend on the variables $x$ and $Q$.
Within the operator product expansion \cite{Wilson:1969zs} 
the structure functions are given by the sum of contributions 
coming from operators of different twists. For the unpolarized 
lepton scattering the even twists larger or equal to two contribute only.
Thus with the account of the twist-4 contribution 
\begin{equation}
F_{2,\rm L}(x,Q)=F_{2,\rm L}^{\rm LT,TMC}(x,Q)
+H_{2,\rm L}(x)\frac{1~{\rm GeV}^2}{Q^2},
\label{eqn:addht}
\end{equation}
where $F_{2,\rm L}^{\rm LT,TMC}$ gives the leading twist (LT) 
with the account of target mass corrections, as calculated in 
Ref.~\cite{Georgi:1976ve}:
\begin{equation}
F_2^{\rm LT,TMC}(x,Q)=\frac{x^2}{\tau^{3/2}}
\frac{F_2^{\rm LT}(\xi_{\rm TMC},Q)}{\xi_{\rm TMC}^2}+
6\frac{M^2}{Q^2}\frac{x^3}{\tau^2}I_2,
\label{eqn:f2tmc}
\end{equation}
\begin{equation}
F_{\rm L}^{\rm LT,TMC}(x,Q)=F_{\rm L}^{\rm LT}(x,Q)
+\frac{x^2}{\tau^{3/2}}(1-\tau)
\frac{F_2^{\rm LT}(\xi_{\rm TMC},Q)}{\xi_{\rm TMC}^2}
+\frac{M^2}{Q^2}\frac{x^3}{\tau^2}(6-2\tau)I_2,
\label{eqn:fltmc}
\end{equation}
where
$$
I_2=\int^{1}_{\xi_{\rm TMC}}dz\frac{F_2^{\rm LT}(z,Q)}{z^2},
$$
\begin{displaymath}
\xi_{\rm TMC}=\frac{2x}{1+\sqrt{\tau}},~~~~\tau=1+\frac{4M^2x^2}{Q^2},
\end{displaymath}
and $F_{2,\rm L}^{\rm LT}$ are the structure functions of twist 2.
Such approach allows us to separate pure kinematical corrections, 
so that the functions $H_{2,\rm L}(x)$ correspond to 
``genuine'' or ``dynamical'' contribution of the twist 4 operators.
Note, that the parametrization (\ref{eqn:addht})
implies, that the anomalous dimensions of the twist 4 operators 
are equal to zero, that is invalid in general case.
Moreover, there are attempts to estimate these anomalous dimensions
from the account of the correlations between partons
(see Ref.\cite{Bukhvostov:1983te}). Meanwhile, in view of 
limited precision of the data, approximation (\ref{eqn:addht})
is rather good (see also discussion in Ref.\cite{Alekhin:2000iq}).

The leading twist structure functions can be expressed in 
factorized form as the Mellin convolution of PDFs $q$ with the coefficient 
functions $C$: 
\begin{equation}
F_{2,\rm L}^{\rm LT}(x,Q)=
\sum_i\int_x^1\frac{dz}{z}
C^i_{2,\rm L}\left[z,\alpha_{\rm s}(\mu_{\rm R}),
Q/\mu_{\rm F}\right]q_i(x/z,\mu_{\rm F}),
\label{eqn:factor}
\end{equation}
where index $i$ marks the partons species and 
$\alpha_{\rm s}$ is running strong coupling constant.
The dependence of PDFs on $Q$ is described by the DGLAP
evolution equations \cite{Gribov:1972ri}
\begin{equation}
Q\frac{\partial q_i(x,Q)}{\partial Q}
=\sum_j\int_x^1 \frac{dz}{z}
P_{ij}\left[z,\alpha_{\rm s}(\mu_{\rm R}),Q/\mu_{\rm R}\right]q_j(x/z,Q),
\label{eqn:dglap}
\end{equation}
and the PDFs evolution is governed by the splitting functions 
$P_{ij}$, which in turn depend on $\alpha_{\rm s}$.
The quantities $\mu_{\rm F}$ and $\mu_{\rm R}$
in Eqs.(\ref{eqn:factor}) and (\ref{eqn:dglap}) give
the factorization and  renormalization scales respectively.
In the $\overline{\rm MS}$ renormalization-factorization scheme,
used in our analysis, these scales are chosen 
equal to the value of $Q$ usually. The splitting and coefficient functions 
can be calculated in perturbative QCD as series in $\alpha_{\rm s}$.
For the coefficient functions these series are completely calculated 
up to the next-to-next-to-leading order (NNLO) \cite{SanchezGuillen:1991iq}; 
for the splitting functions the next-to-leading order (NLO)
corrections are known, while for the NNLO corrections  
a limited set of the Mellin moments \cite{Larin:1997wd}, as well 
as some asymptotes, are available only (see  
references in \cite{vanNeerven:2000ca}).
Nevertheless, there are attempts to analyse the DIS data in the NNLO QCD 
approximation with the consideration of the available moments only  
\cite{Kataev:1998nc,Kataev:1998ce,Santiago:1999pr}, or 
modelling splitting functions \cite{Vogt:1999ik}.
Our analysis is performed in the NLO QCD approximation
with the use of the splitting and 
coefficient functions in $x$-space as they are given in Ref.
\cite{Furmanski:1982cw}. 

The dependence of $\alpha_{\rm s}$ on $Q$ is given by the 
renormalization group equation, which in the NLO QCD approximation
reads\footnote{Analogous equations given in Refs.
\protect\cite{Alekhin:1999za,Alekhin:1999hy} contain misprints, 
meanwhile, the calculations were performed using the correct 
Eqn.(\protect\ref{eqn:alphanlo}).}:
\begin{equation}
\frac{1}{\alpha_{\rm s}(Q)}-\frac{1}{\alpha_{\rm s}(M_{\rm Z})}=
\frac{\beta_0}{2\pi}\ln\left(\frac{Q}{M_{\rm Z}}\right)+
\beta\ln\left[\frac{\beta+1/\alpha_{\rm s}(Q)}{\beta+1/
\alpha_{\rm s}(M_{\rm Z})}\right],
\label{eqn:alphanlo}
\end{equation}
where $\beta=\frac{\beta_1}{4\pi\beta_0}$, 
$\beta_0$ and $\beta_1$ are regular coefficients of 
$\beta$-function: 
$\beta_0=11-(2/3)n_{\rm f}$, $\beta_1=102-(38/3)n_{\rm f}$, 
$n_{\rm f}$ is the number of active fermions, which depends on 
$Q$. In our analysis $n_{\rm f}$ was chosen equal to 3 for 
$Q\le m_{\rm c}$, 4 for $m_{\rm c}\le Q\le m_{\rm b}$, and 5 for 
$m_{\rm b}\le Q\le m_{\rm t}$, where 
$m_{\rm c},m_{\rm b},m_{\rm t}$ are masses of the 
$c$-, $b$- and $t$-quarks correspondingly, and
when $n_{\rm f}$ changes, the continuity of 
$\alpha_{\rm s}(Q)$ is kept (see Ref.\cite{Bernreuther:1982sg} 
for argumentation of this approach). The choice 
of the quark mass value as the threshold for    
$n_{\rm f}$ switching is optional. E.g., in the 
analysis of heavy quark contribution to the DIS
sum rules, given in Ref.\cite{Blumlein:1999sh},
this threshold is chosen equal to $6.5m_{\rm c,b,t}$.
Unfortunately any choice cannot be completely justified, while  
the dependence of results on the variation of threshold,
say in the interval from $m_{\rm c,b,t}$ to $6.5m_{\rm c,b,t}$, generates 
one of the sources of theoretical uncertainties inherent to this 
analysis. Since the value of $\alpha_{\rm s}$ depends on the threshold 
position logarithmically, for estimation of this  uncertainty 
we shifted this threshold value to the logarithmic 
centre of this interval, i.e. from $m_{\rm c,b,t}$ to 
$\sqrt{6.5}m_{\rm c,b,t}$. Very often, 
approximate solutions of Eq.(\ref{eqn:alphanlo}), based on the 
expansions of $\alpha_{\rm s}$ in inverse powers of 
$\ln(Q)$ are used in calculation. Inaccuracy of 
these expansions for evolution of $\alpha_{\rm s}$ 
from $O({\rm GeV})$ to $M_{\rm Z}$ may be as large as 
$0.001$ \cite{Barnett:1996hr}, which is comparable with 
the experimental uncertainties of the $\alpha_{\rm s}(M_{\rm Z})$
value extracted from the data. In order to escape these uncertainties we use 
in the analysis the exact numerical solution of Eq.(\ref{eqn:alphanlo}) 
instead.

Since we use the truncated perturbative series, the 
results depend on the factorization scale $\mu_{\rm F}$
and the renormalization scale  
$\mu_{\rm R}$. These dependences cause 
additional theoretical uncertainties of the analysis
The accurate estimate of 
these uncertainties is difficult, because the possible range of 
the scales variation is undefined and besides, one is to 
change factorization scheme as well. In our analysis we 
estimate only the theoretical uncertainty due to the choice 
of $\mu_{\rm R}$ in the evolution equations (\ref{eqn:dglap})
using the approach described in Ref.\cite{Martin:1991jd}. 
In accordance with this approach the renormalization scale $\mu_{\rm R}$ 
is chosen equal to $k_{\rm R}Q$ and the NLO
evolution equations are modified in 
the following way
$$
Q\frac{\partial q_i(x,Q)}{\partial Q}
=\frac{\alpha_{\rm s}(k_{\rm R}Q)}{\pi}\sum_j\int_x^1 \frac{dz}{z}
\left\{P^{(0)}_{ij}(z)+\frac{\alpha_{\rm s}(k_{\rm R}Q)}{2\pi}
\left[P^{(1)}_{ij}(z)+\beta_0 P^{(0)}_{ij}(z)\ln(k_{\rm R})\right]
\right\}q_j(x/z,Q),
$$
where $P^{(0)}$ and $P^{(1)}$ are respectively the LO and NLO 
coefficients of the splitting functions series.
The change of results under variation of 
$k_{\rm R}$ from 1/2 to 2 gives an estimate of the 
error due to renormlization scale uncertainty. 
Evidently that, by definition, 
this uncertainty is connected with the impact of 
unaccounted terms of the perturbative series. 
 
In order to obtain the PDFs from evolution equations, one 
is to supply a boundary conditions at some starting value 
$Q_0$. The $x$-dependence of PDFs cannot be calculated from 
the modern strong interaction theory, it is determined from 
the comparison with data. Usual parametrization of the PDFs at $Q_0$ 
reads 
$$
xq_i(x,Q_0)=x^{a_i}(1-x)^{b_i}.
$$
For this parametrization the behaviour of $q$ at low $x$ is motivated by the Regge
phenomenology (see, e.g., book \cite{INDUR}) and at high 
$x$, by the quark counting rules \cite{Matveev:1973ra,Brodsky:1973kr}. 
If such a simple form is insufficient for the fair
data description, polynomial-like factors are added.
Value of $Q_0$ is arbitrary, but it is natural to choose 
it as $O({\rm GeV})$ to allow for simple interpretation of 
the boundary PDFs. Meanwhile, it was recently shown in 
Ref.\cite{Alekhin:1999kt}, that the choice of $Q_0$ 
is important to provide the results stable with respect 
to the account of higher order QCD corrections
(see also Ref.\cite{Kataev:1998ce}). At low $Q_0$ the twist 4 contribution
extracted from the data is less sensitive to 
the choice of the renormalization scale $\mu_{\rm R}$ 
in Eq. (\ref{eqn:dglap}), than at high $Q_0$. 
The $\alpha_{\rm s}(M_{\rm Z})$ behaves in opposite way,
and then the choice $Q_0^2=9~{\rm GeV}^2$, made in our 
analysis, provides stability of the $\alpha_s$ and PDFs values
if the NNLO QCD corrections are considered.  

\begin{figure}
\centerline{\epsfig{file=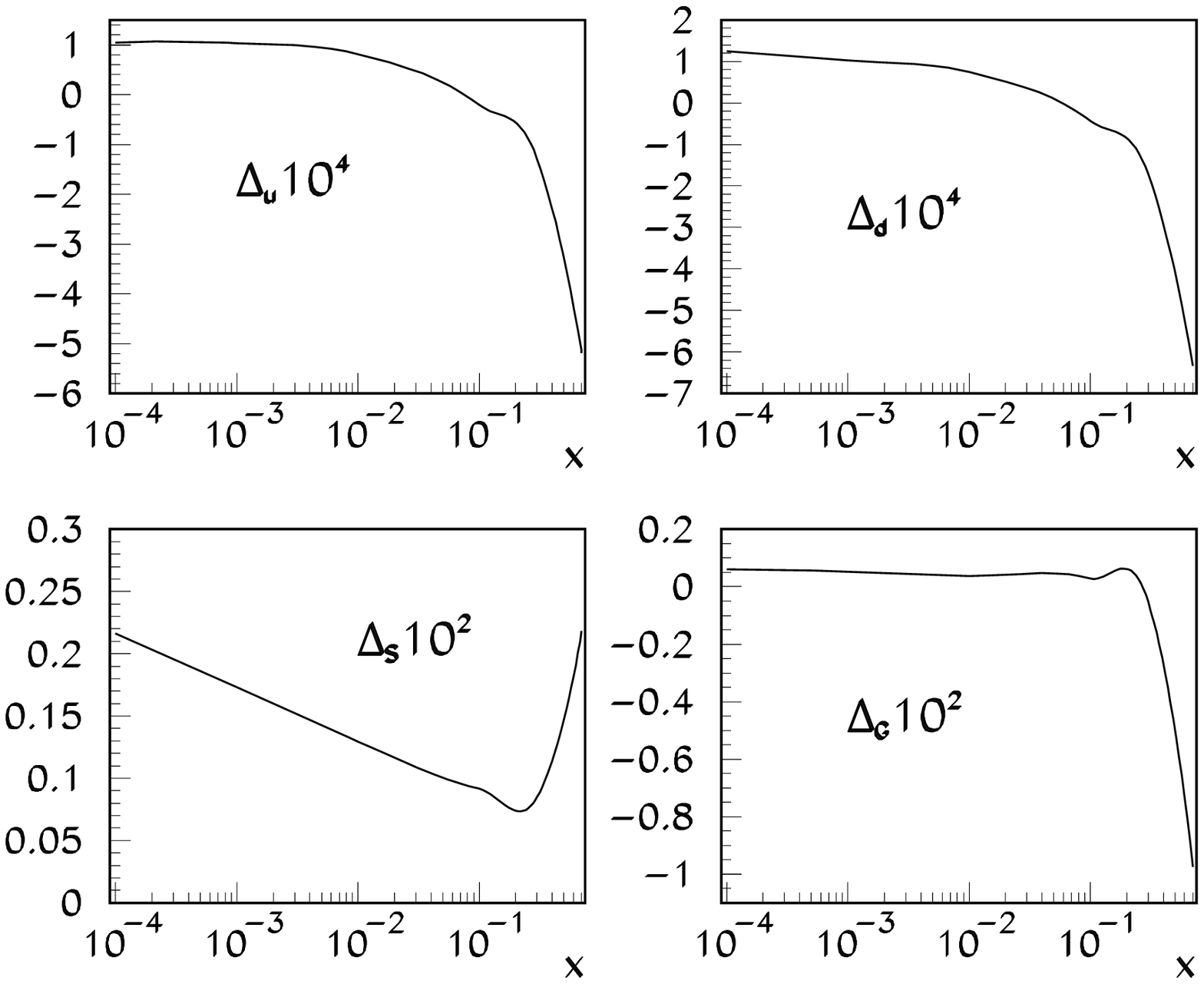,width=12cm,height=10cm}}
\caption{$\Delta$ is relative precision of our code used for 
the evolution equations integration. Indices $u$ and $d$ 
correspond to the valence quarks;
$S$, to the sea quarks; $G$, to gluon.}
\label{fig:bench}
\end{figure}

Despite of the fact, that the evolution equations have been  
used in the DIS data analysis
for many years, no unique approach for solving them exists.
Analytical expressions can be obtained only for the simplified 
splitting functions, and direct numerical approaches demand 
threefold integration, which is time consuming.
There are semi-analytical approaches, based on expansion of 
PDFs in terms of selected sets of functions, but such approaches, 
as a rule, lead to loosing of the universality with respect 
to the choice of splitting functions and require 
careful control of the calculations precision. Due to the form 
of the evolution equation kernel is rather complicated, 
correct implementation of a sophisticated integration algorithm 
meets the difficulties.
In the comparative analysis of different codes, used for the 
DGLAP equations integration,
the codes of the  CTEQ and MRST collaborations were found to 
contain the bugs (see Ref.\cite{Blumlein:1996rp}).
Taking into account these points, we use in the analysis 
our own code for direct numerical integration of 
Eq.(\ref{eqn:dglap}), based on the Euler predictor-corrector algorithm
(see Ref.\cite{MATHBOOK}). This code allows one to modify kernels 
of the evolution equations in order to debug the code, to control 
the calculation precision, to take into account effects of new physics, 
and to implement special cases of evolution.
Integration region can be expanded easily, and the integration 
precision is regulated by the external parameters of the code.
For typical values of these parameters
the code integration precision, as estimated 
using benchmark described in Ref.\cite{Blumlein:1996rp}, 
is given in Fig.~\ref{fig:bench}. One can see that the relative 
precision is better, than 0.001 in the region $x\lesssim0.5$ and 
better, than 0.01 in the region $x\gtrsim0.5$.
This is well enough for our purposes, since 
the errors on data are larger, than the integration 
errors for all $x$.

Since Eq.(\ref{eqn:dglap}) is valid for 
massless partons only, the heavy quarks contribution, 
which is significant at low $x$, should be considered
in a peculiar way. In 
the approach described in Ref.~\cite{Collins:1986mp} 
the heavy quarks are considered as massless ones.
They are included 
into the general evolution starting from a threshold 
value of $Q$, which is proportional to the quark mass, while at 
the values of $Q$ lower the threshold these distributions 
are put to zero. Evidently, in this approach the heavy quarks
contribution is overestimated in the vicinity of the threshold. 
Alternative way to consider the heavy quarks contribution is to 
calculate it using the photon-gluon fusion model
of Ref.\cite{Witten:1976bh}. At high $Q$ and low $x$
``large logarithms'' arise in the elementary cross section of this 
process, that may demand its resummation \cite{Shifman:1978yb}.
Meanwhile, as it was shown in Ref.\cite{Gluck:1994dp}, 
the region of $x$ and $Q$, where the resummation is really needed 
lays outside the region of the available DIS data. For this reason
we calculated the $c$- and $b$- quarks contributions 
to the structure functions $F_{2,\rm L}$ using the 
photon-gluon fusion model with the NLO coefficient functions 
of Ref.\cite{Laenen:1993xs} and the 
renormalization/factorization scales equal to 
$\sqrt{Q^2+4m^2_{\rm c,b}}$ at the quark masses 
$m_{\rm c}=1.5~{\rm GeV}$ and $m_{\rm b}=4.5~{\rm GeV}$.

The LT contribution to the DIS structure functions 
is rather well understood both from theoretical and experimental 
points of view. Since this contribution depends on $Q$ weakly
one can reject the low $Q$ data points and leave  
the data set, which is both statistically significant, and 
can be analysed within perturbative QCD in order to determine 
the LT $x$-dependence. The HT contribution is worse known, 
than the LT one. The theoretical analysis of the HT $x$-dependence 
is equally difficult as for the LT $x$-dependence and, as a 
result, it should be determined from data. Meanwhile, due to the fast 
fall of the HT contribution with $Q$ it is significant for
$Q^2\lesssim10~{\rm GeV}^2$ only. At very low $Q$ 
the subtraction of the LT contribution, as calculated 
in perturbative QCD is problematic due to the rise 
of $\alpha_{\rm s}$. As a result, only the data for 
$Q^2\lesssim 1~{\rm GeV}^2$ $Q$ can be used for the HT extraction 
and the results precision is poor.

Study of the possibility to separate the HT and LT contributions
has a long history (see Refs. 
\cite{Berger:1979du,Abbott:1980as,Buras:1980yt,Bednyakov:1984gh}). 
Despite of that the $Q$-dependences of these contributions are different, 
in the limited range of $Q$ the HT power corrections
can simulate the logarithmic LT behaviour
\cite{Penin:1997zk}. Moreover, as it was shown in 
Refs.\cite{Abbott:1980as,Mahapatra:1997av},
the power corrections can almost entirely describe the 
scaling violation observed for the DIS data,
if the data precision is limited. In particular, this 
causes large correlations between the fitted values of 
$\alpha_{\rm s}$ and the HT contribution.
This correlation leads to the rise of the fitted parameters errors. 
The rise of errors is unpleasant effect, moreover,  
the fitted model non-linearity can become essential as a result. 
Besides, the fit results become less stable with respect to 
the change of the non-fitted parameters and adoptions of the fitted model, 
i.e., the theoretical errors on the fitted parameters rise also.
Finally, if large correlations between parameters occur, 
the second derivative matrix for the minimized functional
is poor determined and 
the calculations inaccuracies increase when its inversion. 
For this reason in order to get satisfactory precision of 
the parameters errors one is to guarantee better precision 
of the fitted model calculation, which may be time consuming, 
if manifold integration is involved. Due to this is the case 
for our analysis,  
estimation of the correlation coefficients between the fitted 
values of 
$\alpha_{\rm s}$ and the HT contribution is non-trivial problem.

\section{Data used in the fit and starting PDFs}

We fit the PDFs to the data on the charged leptons DIS off 
proton and deuterium given in Refs.
\cite{Whitlow:1992uw,Benvenuti:1989rh,Arneodo:1997qe,Adams:1996gu,Aid:1996au,Derrick:1996hn}.
The data points with $Q^2<2.5~{\rm GeV}^2$ were not used 
in the analysis in order to  
reject the region, where $\alpha_{\rm s}$ is rather large 
and the NNLO order QCD correction may be important. The points with 
$x>0.75$, for which the nuclear corrections are large,
were removed also. The data used in the analysis occupy the region 
$10^{-4}\lesssim x\le 0.75$, 
$2.5~{\rm GeV}^2\le Q^2\lesssim 5000~{\rm GeV}^2$.
The number of data points for each experiment is given in  
Table~\ref{tab:global}.

The starting PDFs were initially parametrized at $Q_0^2=9~{\rm GeV}^2$ 
as follows: 
\begin{equation}
xq_i(x,Q_0)=A_ix^{a_i}(1-x)^{b_i}(1+\gamma^i_1\sqrt{x}+\gamma^i_2 x)
\label{GENPDF}
\end{equation}
and then the parameters $\gamma$, which agree with zero within errors,
were by turn fixed at zero till such 
parameters existed. Evidently, the fit quality could not get 
worse, when such parameters are fixed. The PDFs functional 
form resulted from this simplification 
and used in the final fit reads:
\begin{displaymath}
xu_{\rm V}(x,Q_0)=\frac{2}{N^{\rm V}_{\rm u}}
x^{a_{\rm u}}(1-x)^{b_{\rm u}}(1+\gamma_2^{\rm u}x),
\end{displaymath}
\begin{displaymath}
xu_{\rm S}(x,Q_0)=\frac{A_{\rm S}}{N_{\rm S}}
\eta_{\rm u} x^{a_{\rm su}}(1-x)^{b_{\rm su}},
\end{displaymath}
\begin{displaymath}
xd_{\rm V}(x,Q_0)=\frac{1}{N^{\rm V}_{\rm d}}x^{a_{\rm d}}(1-x)^{b_{\rm d}},
\end{displaymath}
\begin{displaymath}
xd_{\rm S}(x,Q_0)=\frac{A_{\rm S}}{N^{\rm S}}x^{a_{\rm sd}}(1-x)^{b_{\rm sd}},
\end{displaymath}
\begin{displaymath}
xs_{\rm S}(x,Q_0)=\frac{A_{\rm S}}{N^{\rm S}}\eta_{\rm s}
x^{a_{\rm ss}}(1-x)^{b_{\rm ss}},
\end{displaymath}
\begin{displaymath}
xG(x,Q_0)=A_{\rm G}x^{a_{\rm G}}(1-x)^{b_{\rm G}}
(1+\gamma^{\rm G}_1\sqrt{x}+\gamma^{\rm G}_2 x),
\end{displaymath}
where $u,d,s,G$ are the up, down, strange quarks,
and gluons distributions respectively; 
indices $V$ and $S$ correspond to the valence 
and sea quarks. The parameters $N^{\rm V}_{\rm u}, N^{\rm V}_{\rm d}$  and 
$A_{\rm G}$ were not fitted, instead they were calculated 
from the other parameters using the conservation of the partons momentum 
and the fermion number. The parameter $N^{\rm S}$ was calculated using 
the relation 
$$
2\int_0^1x\bigl[u_{\rm s}(x,Q_0)+d_{\rm s}(x,Q_0)
+s_{\rm s}(x,Q_0)\bigr]dx=A_{\rm S}.
$$
It is well known, that the charged leptons data do not allow to 
confine the sea quarks contribution. For this reason the 
parameter $\eta_{\rm s}$ was fixed at 0.42, 
which agrees with the recent results of the NuTeV
collaboration, given in Ref.\cite{Adams:1999sx}.
The other sea distributions parameters
were constrained as $a_{\rm su}=a_{\rm sd}=a_{\rm ss}$, 
$b_{\rm ss}=(b_{\rm su}+b_{\rm sd})/2$.

The DIS cross sections calculated from the QCD evolved PDFs  
using Eq.(\ref{eqn:discs}) with the account of the TMC 
corrections given by Eq.(\ref{eqn:f2tmc}) and the twist 4 
contribution in additive form as in Eq.(\ref{eqn:addht}), 
were fitted to the cross section data\footnote{Since the 
high $Q$ data from the H1 and 
ZEUS experiments were corrected for the 
$Z$-boson contribution, Eq.(\protect\ref{eqn:discs}) 
is applicable for these data also.}. The HT contributions
to the proton and neutron structure functions $F_2$ were 
parametrized by separate functions 
$H_2^{\rm p}$ and $H_2^{\rm n}$ respectively, 
and the HT contributions 
to the proton and neutron structure 
functions $F_{\rm L}$, by the common function  
$H^{\rm N}_{\rm L}$, since the latter coincide within errors.
The functions $H_2^{\rm p,n}$ and $H_{\rm L}^{\rm N}$
were parametrized in the model independent way: 
at $x=0.,0.1,\dots 0.8$ their values were fitted to the data, 
and between these point the functions were linearly interpolated.  
The common approach for the PDFs global fits is to use 
data on $F_2$ instead of the data on cross sections.
Within this approach one ignores the fact,  
that the $F_2$ values given by different experiments  
are often extracted from the cross sections using 
different values of $F_{\rm L}$.
In our fit the $F_{\rm L}$ contribution to the cross 
section was calculated iteratively and, efficiently, the data were 
reduced to the common value of $F_{\rm L}$. 
Since the $F_{\rm L}$ contribution rises with $y$, 
the effect of this reduction is 
more important at high $y$. Due to the collision energy of 
each experiment is limited, the highest values of $y$ 
correspond to the minimal values of $x$.
For this reason 
the $F_2$ data points shifts due to the reduction to the common 
value of $F_{\rm L}$ are not very significant in average, but 
at the edges of the experiments data regions may reach several 
percents. Note, that at low $x$ the $F_{\rm L}$ value 
strongly depends on the gluon distribution and, hence, in the fit 
to the cross sections data an additional constraint for 
the gluon distribution occurs, i.e. it 
is better confined, as compared to the fit to the $F_2$ data.

The TMC correction is most important for the 
SLAC data, less important for the BCDMS data, almost 
unimportant for the NMC data, and negligible for the others.
Note, that our TMC correction to $F_2$
given by Eq.(\ref{eqn:f2tmc})
differs from that applied in Ref.\cite{Martin:1995kk}, where
the substitution 
\begin{equation}
F_2^{\rm LT,TMC}(x,Q)=F_2^{\rm LT}(\xi_{\rm TMC},Q)
\label{eqn:tmcmrs}
\end{equation}
was used to account for the target mass effect.
The numerical difference between these two approaches is 
maximal at high $x$ and low $Q$, e.g., for the SLAC data 
it reaches 40\%. Besides, our TMC correction, contrary 
to that given by Eq.(\ref{eqn:tmcmrs}), changes sign at 
$x\approx0.5$.

The deuterium data were corrected for the Fermi motion effect
as in the model of Ref.\cite{Atwood:1973zp} with 
the deuterium wave function from Ref.\cite{Lacombe:1980dr}. 
The deuterium correction value rises with $x$ and reaches 16\% 
for the SLAC data. This correction was calculated 
iteratively in the fit to provide consistency of the 
analysis. The two-dimensional integrals involved in the model   
were calculated using the code of Ref.\cite{Sokolov:1988mw},
which provides better numerical stability, than the standard
codes based on the Gauss integration algorithm.
For the calculations time saving we adopted, that 
the Fermi motion correction for the structure function 
$xF_1=F_2-F_{\rm L}$ is the same, as for the structure 
function $F_2$ (we checked that this adoption does not 
significantly affect the results).

\section{Fitting procedure and results}

The fitted parameters including the PDFs parameters, the value
of $\alpha_{\rm s}$, and the coefficients of the functions 
$H_{2,\rm L}$ were determined from the minimization of 
the functional
\begin{equation}
\chi^2=\sum_{K,i,j}(f_i-\xi_Ky_i)E_{ij}(f_j-\xi_Ky_j),
\label{eqn:chi2}
\end{equation}
where $E_{ij}$ is inverse of the covariance matrix
$C_{ij}$,
$$
C_{ij}=\xi_K^2\delta_{ij}\sigma_i\sigma_j
+f_if_j(\vec{\eta}_i^K \cdot \vec{\eta}_j^K),
$$
index $K$ runs through the data subsets corresponding to the 
different experiments and the different targets within one experiment,
indices $i,j$ run through the data points in these subsets.
The other notations: $y_i$ are the measurements;
$\sigma_i$ are the statistical errors; 
$\xi_K$ are the renormalization factors;
$f_i$ are the fitted model calculations depending on the fitted 
parameters; $\vec\eta^{K}_i$ are the systematic errors vectors 
(the dimensions of these vectors for each experiment are given 
in Table~\ref{tab:global} as NSE). The systematic errors were considered as 
multiplicative, that is natural way for the counting experiments. 
All systematic errors, excluding the normalization errors on the old SLAC 
experiments, were accounted for in the covariance matrix. 
The data from the old SLAC experiments, as they were given in 
Ref.\cite{Whitlow:1992uw}, are the result of re-analysis
of the original experimental data published earlier 
(for the details see Ref.\cite{Whitlow:1990dr}).
One of the purposes of this re-analysis was to renormalize 
the old data on the data from dedicated experiment SLAC-E-140.
However, due to the latter did not release the proton target data, 
the renormalization of the proton data was performed using 
the experiment SLAC-E-49B as a ``bridge''.
Such technique certainly brings additional uncertainties on the re-analysed data.
In order to escape those uncertainties we performed the independent renormalization of 
the old SLAC experiments without a ``bridging'', that is possible in our case,
since we use more proton data, than in the analysis of Ref.\cite{Whitlow:1992uw}.
For this purpose we fitted the factors $\xi_K$ for each 
target of each old SLAC experiment independently.
Alongside, to keep the analysis consistency the errors due to normalization 
uncertainties of the old SLAC experiments, 
given in Ref.\cite{Whitlow:1992uw}, were cancelled out.
For other experiments the parameters $\xi_K$ were fixed at 1.
The asymmetrical systematic errors on the ZEUS data were 
symmetrized, when including in the covariance matrix, 
and systematic errors on the BCDMS data for the proton and deuterium targets
were considered as perfectly correlated.

\begin{figure}
\centerline{\epsfig{file=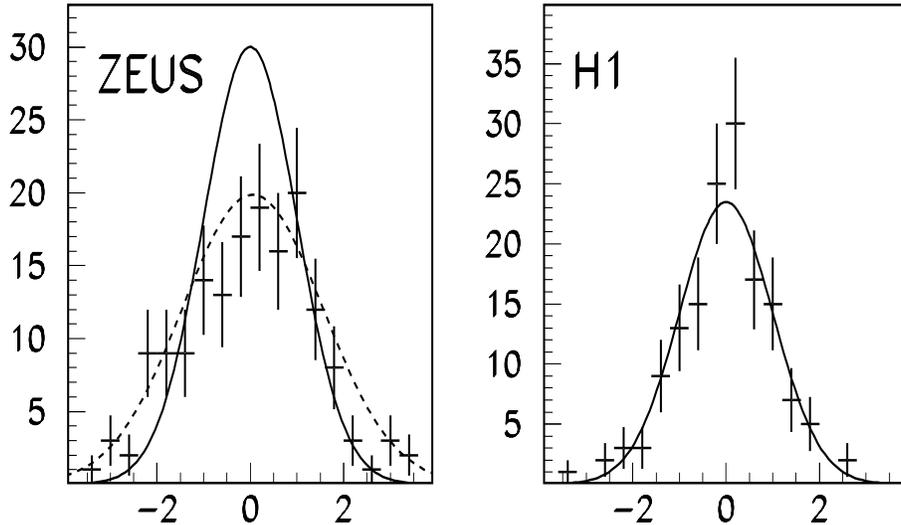,width=12cm,height=7cm}}
\caption{The distribution of diagonalized residuals for the 
ZEUS and H1 data (full curves: normal distribution, 
dashes: the Gauss distribution with the dispersion and the mean 
equal to the dispersion and the mean of the residuals distribution).
All curves are normalized to the number of points in each experiment.}
\label{fig:resid}
\end{figure}

The statistical properties of the estimator based on covariance matrix (CME)
were considered in Ref.\cite{Alekhin:2000es} 
in comparison with the statistical properties of the simplest 
$\chi^2$ estimator (SCE), based on the minimization of the functional 
$$
\chi^2=\sum_{i} \frac{(f_i-y_i)^2}{\sigma_i^2},
$$
which is often used in particle physics for the analysis 
of data including the correlated ones as well. For the CME 
the fitted parameters systematic errors 
due to the data systematic errors are automatically
included in the total error; 
for SCE the parameters systematic errors  
are estimated as the shift  
of the parameters under the shift of the data by the value of 
their systematic errors. The SCE dispersion is always larger, than 
the CME dispersion and, as it was shown in Ref.\cite{Alekhin:2000es}, 
the ratio of these dispersions can reach several units
for realistic cases. It was shown also, that 
the CME is unbiased if the systematic errors 
on the parameters are not much more, than the statistical ones.
In order to control the estimator bias one can trace 
the value of the net residual $R$, equal to the mean of 
weighted residual $(f-y)/\sqrt{\sigma^2+(f\eta)^2}$ .
The $\chi^2$ values and the net residuals for 
the total data set and for each 
experiment separately calculated at the parameters values
fitted using the CME are given in Table~\ref{tab:global}.
One can see, that the net residual value is within
its standard deviation\footnote{The $R$ standard deviation was calculated using 
Eq.~(3.11) from Ref.\protect\cite{Alekhin:2000es}.} 
and the data description is good, 
excluding description of the ZEUS data.
For more detailed analysis of the confidence of the 
ZEUS data description we calculated for those data 
the diagonalized residuals $r^{\rm D}$ using the relation
$$
r^{\rm D}_i=\sum^N_{j=1}\sqrt{E}_{ij}(f_j-y_j),
$$
where indices $i,j$ run through the data points. If data are 
well described by fitted model, then for large $N$ the values
of $r^{\rm D}_i$ obey the normal distribution, i.e. 
the Gauss distribution with zero mean and the dispersion
equal to 1.
The distribution of $r^{\rm D}_i$ for the ZEUS data is given in 
Fig.~\ref{fig:resid}. Evidently it does not agree with 
the normal distribution, that is not surprising, since the 
data description is poor. Note, that the diagonalized residuals
mean is small for the ZEUS data (0.05), meanwhile, 
the dispersion is equal to 2.1, 
i.e. it is far from the normal distribution dispersion. It is 
difficult to ascribe this discrepancy to the shortcoming of the 
fitted model, since, as it seen from Fig.~\ref{fig:resid},
analogues distribution for the H1 data agrees with the 
normal distribution perfectly, whereas both experiments gained similar 
statistical and kinematical coverage. One more possible 
explanation of this disagreement is, 
that the systematic errors given by the ZEUS collaboration 
are underestimated, but still are Gaussian distributed.
In such cases the PDG scales the errors so that 
$\chi^2/{\rm NDP}$ becomes equal to 1 (see review \cite{Barnett:1996hr}).
In our case this approach cannot be used, since number 
of independent sources of the systematic errors in the ZEUS experiment 
is large and a lot of variants of such rescaling can be applied.
Besides, the distribution of residuals would
remain non-Gaussian after the errors rescaling 
(see dashed curve in Fig.~\ref{fig:resid}).
Driven by this consideration one can suppose that systematic errors
on the ZEUS data are non-Gaussian distributed (but with zero mean)
and then $\chi^2/{\rm NDP}$ must not be equal to 1.
If so, the fitted parameters, which are confined by the ZEUS data, 
also may be distributed in arbitrary way (see in this connection 
Ref.\cite{Giele:1998gw}). Due to exact estimation of 
the confidence intervals for unknown distribution is impossible, 
we recommend for this purpose, in particular for 
evaluating the PDFs errors at low $x$,
the robust estimate of the confidence intervals, based on the Chebyshev 
inequality (see discussion in Ref.\cite{Alekhin:2000es}).
 
The dispersion of the net residual $R$ is maximal for the
SLAC-E-140, BCDMS, and NMC data sets. (Remind, that this dispersion 
rises with the increase of data correlation, full correlation 
corresponds to the dispersion of $R$ equal to 1).
Thus, one can conclude, that  
the account of the BCDMS and NMC data correlations
has the largest impact on the analysis results, 
since number of points in the 
SLAC-E-140 data set is small. This conclusion 
is in line with the results of Ref.\cite{Alekhin:1999hy},
where it was obtained, that in the combined fit to 
the non-singlet SLAC-BCDMS data the 
account of the BCDMS data correlations leads to much more 
significant shift of the parameters, than the account of the 
SLAC data correlations. The value of $R$ for the total data set 
is well within its standard deviation, that confirms
the fit unbiasness.

The fitted PDFs parameters are given in Table~\ref{tab:pdfpars}. 
We underline, that in our fit the universality of the 
valence $u$- and $d$-quarks
behaviour at low $x$ is not initially assumed, contrary to the   
popular global fits practice, and the fit results confirm this 
universality with the few percents precision.
At the same time the 
Regge phenomenology prediction (see, e.g., book~\cite{INDUR})
\begin{equation}
a_{\rm u}=a_{\rm d}=0.5
\label{eqn:redge}
\end{equation}
is in disagreement with the fit results\footnote{We especially 
underline this point, since Eq.(\protect\ref{eqn:redge})
is often used for theoretical estimates.}.
A possible interpretation of this disagreement is, that 
Eq.(\ref{eqn:redge}), as it is deduced, is 
not related to a particular value of $Q$, while 
the QCD evolution does change the PDFs $x$-behaviour.
As it was shown in Ref.\cite{Gross:1974fm}, for the non-singlet 
distributions at low $x$ this change is not very significant, 
but at least partially it can help to explain the observed 
disagreements. The values of the parameters $a_{\rm u}$ and $a_{\rm d}$
agrees with the results of our earlier analysis of 
Ref.\cite{Alekhin:1999za} and with the value of the parameter
describing the low $x$-behaviour of the non-singlet neutrino
structure function $xF_3$, which was obtained from the fit to 
the CCFR data in Refs.~\cite{Seligman:1997fe,Kataev:2000dp}.
For the obtained values of the 
parameters, which describe the valence $u$- and $d$-quarks
behaviour at high $x$, the relation 
$b_{\rm d}=a_{\rm u}+1$ holds with good precision, 
in line with the quark counting rules. Meanwhile, the absolute 
values of these parameters deviate from the 
quark counting rules predictions $b_{\rm u}=3$, $b_{\rm d}=4$.
This disagreement can also be due to the QCD evolution, moreover
for the non-singlet distribution 
the evolution effect is stronger at high $x$.

As one can see from Table~\ref{tab:pdfpars}, the 
systematic errors on the parameters describing 
the valence $u$-quark distributions
at high $x$ and the sea quarks distributions at 
low $x$ are especially large. At the same time  
the ratio of the total error to the pure statistical one is 
$O(1)$ for any fitted parameter, that 
guarantees their unbiasness. In order to estimate the sensitivity 
of the parameters values to the approach used for the account of 
the systematic errors we performed the fit using the SCE
and the fit with the statistical and systematic error combined 
in quadrature. Results of these fits are also given in 
Table~\ref{tab:pdfpars}. One can see, that in the SCE fit 
the central values of some parameters are shifted by more, than 
two standard deviations, as compared with the CME fit and 
the shift is larger for the parameters with large 
ratio of the systematic errors to the statistical ones, 
in particular, for $b_{\rm u}$ and $\gamma^2_{\rm u}$.
Nevertheless, the SCE fit provides correct estimate 
of the parameters, the only shortcoming of the SCE
is that the SCE errors  
may be several times larger, that the CME errors. 
In our analysis maximal ratio of these errors is about 5 and 
within the errors the results of both fits agree.
At the same time the fit with the statistical and 
systematic errors combined in quadrature does may 
give incorrect estimate of the parameters, since the 
data correlation information is lost in this case.
As a result, the central values of some parameters,
in particular, $b_{\rm G}$ and $a_{\rm sd}$, obtained 
from this fit are shifted from the CME fit results by the 
statistically significant values (see Table~\ref{tab:pdfpars}).
Some parameters errors obtained in these two fits 
are very different also, e.g., the errors on 
$\alpha_{\rm s}(M_{\rm Z})$ 
and the parameters describing the gluon distribution at high $x$.
These differences evidently may lead to the fake  
disagreements with another experimental results
and cause discussions on new physics manifestation, if 
the results of the fit performed without the account of the data 
correlations are used for the comparison. (The example of 
resolution of such 
``disagreement'' encountered in the comparison of the 
SLAC-BCDMS and LEP data
on $\alpha_{\rm s}$ was given in Ref.\cite{Alekhin:1999hy}).

\section{The experimental PDFs uncertainties}

\begin{figure}
\centerline{\epsfig{file=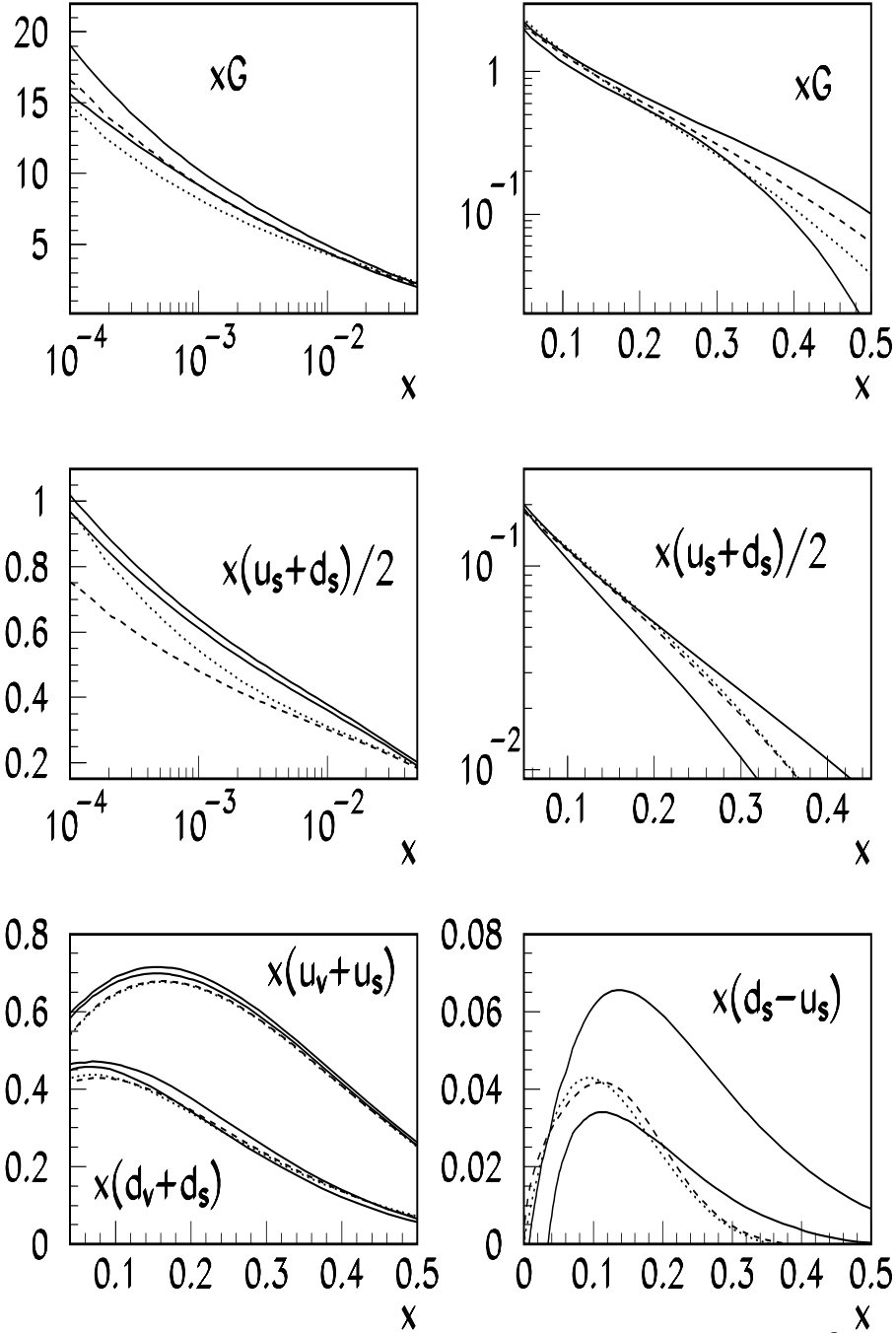,width=12cm,height=18cm}}
\caption{The 1$\sigma$ experimental error bands for our 
PDFs at $Q^2=9~{\rm GeV}^2$ (full lines).
For comparison the nominal 
MRST99 (dots) and CTEQ5 (dashes) PDFs are also given.}
\label{fig:pdfs}
\end{figure}

\begin{figure}
\centerline{\epsfig{file=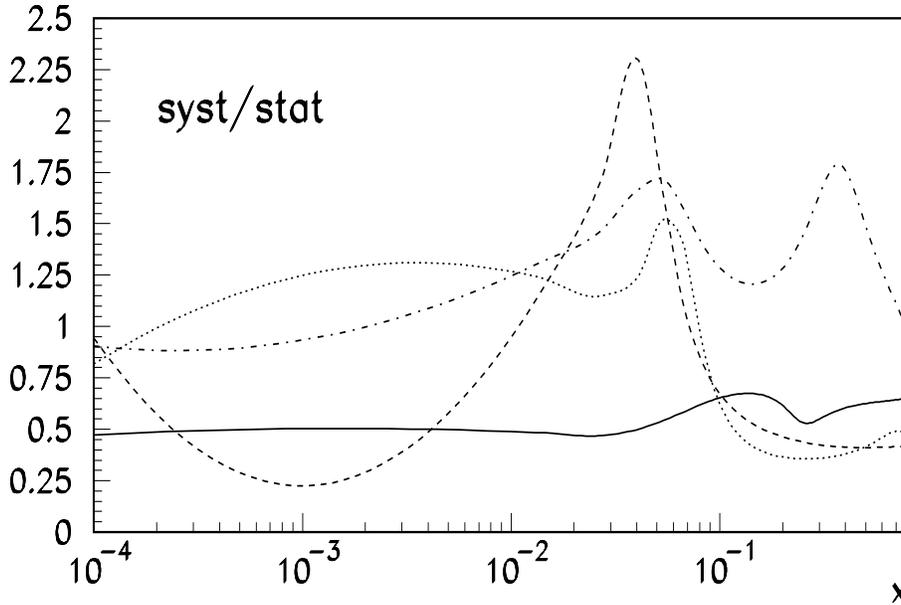,width=12cm,height=8cm}}
\caption{The ratio of the systematic errors on the 
fitted PDFs to the statistical ones (full lines: the gluon distribution;
dashes: the total sea one; dots: the $d$-quark one; dotted-dashes: the $u$-quark one).} 
\label{fig:pdfs_syst}
\end{figure}

\begin{figure}
\centerline{\epsfig{file=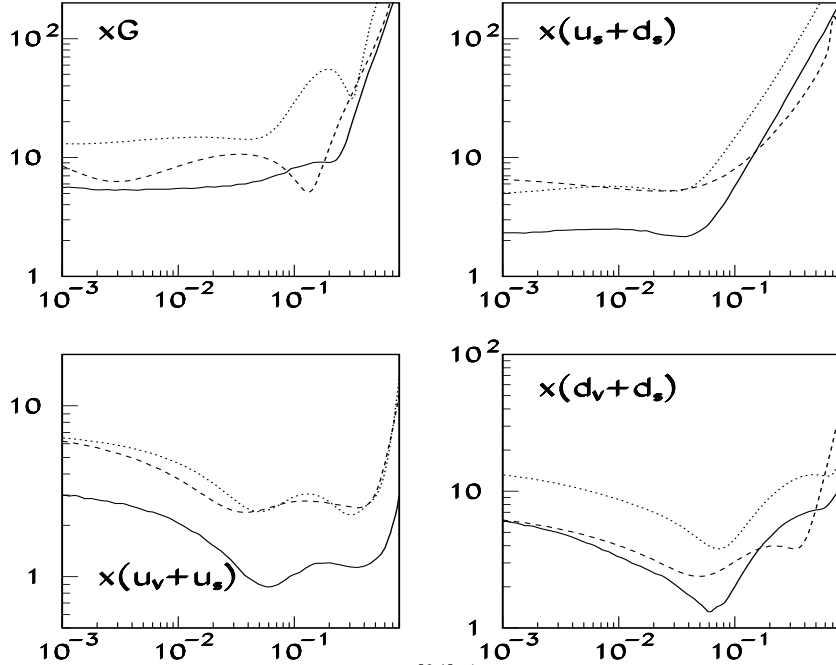,width=12cm,height=9cm}}
\caption{The relative experimental PDFs errors [\%]
(full lines: our analysis, dashes: the analysis of Ref.\protect\cite{Botje:2000dj}).
For comparison the relative experimental PDF errors obtained in our analysis 
from the SCE fit are also given (dots).}
\label{fig:bot}
\end{figure}

\begin{figure}
\centerline{\epsfig{file=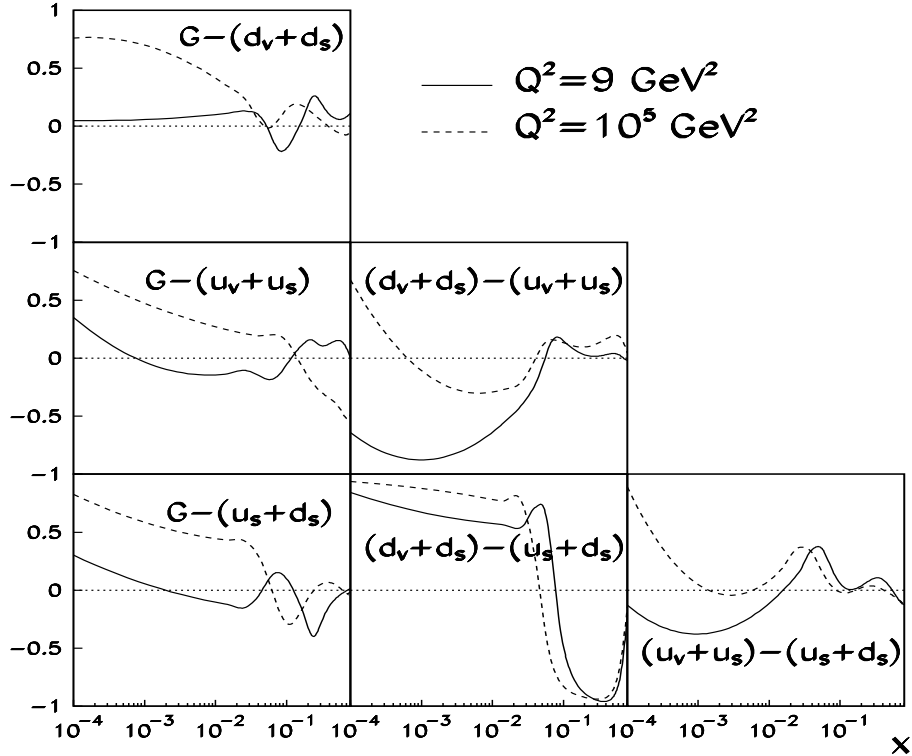,width=12cm,height=10cm}}
\caption{The PDFs correlation coefficients at different $Q^2$.}
\label{fig:pdfs_corr}
\end{figure}

The fitted PDFs with their experimental errors, 
including both statistical and systematic ones 
are given in Fig.~\ref{fig:pdfs}, 
and the relative experimental errors on the PDFs 
are given in Fig.~\ref{fig:pdfs_err}. To estimate the separate  
contribution of the systematic errors to the total ones
we calculated the parameters dispersions  
keeping the central values of the 
fitted parameters, but without the account of systematic errors on the data.
Then we extracted these reduced dispersions from the total 
dispersions of the parameters and took the square roots of these differences 
as the systematic errors on the parameters.
The ratio of systematic errors on the selected PDFs to their statistical errors 
is given in Fig.~\ref{fig:pdfs_syst}. As it was noted above, 
the systematic errors impact is more important for the 
$u$-quark distribution at all $x$ in question and for 
$d$-quark distribution at low $x$. The PDFs errors, as well as their 
parameters errors, depend on the approach used for the accounting 
of systematic errors. The PDFs errors
obtained in the CME and SCE fits are compared in Fig.~\ref{fig:bot} 
and one can see, that for the latter the errors
are several times larger generally.
The errors on PDFs obtained from the CME fit 
in our earlier analysis of Ref.\cite{Alekhin:1999za} 
are also given in the same figure. In that analysis we used data of 
Refs.\cite{Benvenuti:1989rh,Arneodo:1997qe,Aid:1996au,Derrick:1996hn}
with $Q^2>9~{\rm GeV}^2$ and $W>4~{\rm GeV}$.
At small and moderate $x$
the errors on PDFs obtained in the earlier 
analysis are several times larger, that the PDFs errors obtained in 
the present analysis.
At high $x$ these errors are of the same order, 
and for some PDFs the earlier analysis errors are even smaller.
This occurs due to in the analysis of Ref.\cite{Alekhin:1999za} 
the HT contribution was fixed at zero, that decreased 
the PDFs errors. The correlation coefficients 
matrix for the PDFs parameters is given in Table~\ref{tab:pdfcor}
and the selected PDFs correlation coefficients are given in 
Fig.~\ref{fig:pdfs_corr}. The correlations are larger for 
the valence and sea quarks distributions. This can be readily 
understood, since these distributions contribute to the 
charged leptons DIS structure 
functions as the sum and hence can be separated hardly.
Due to the large correlations between some PDFs the ratio of the systematic
errors on their linear combinations to the statistical errors 
on these combinations may be not proportional to  
such ratios for the PDFs themselves.
For example, as one can see in Fig.~\ref{fig:pdfs_syst}, 
the relative systematic errors on the sum of non-strange quarks 
distributions at low $x$
are significantly smaller, than for the $u$- and $d$-quarks
distributions separately.

The relative experimental error on the gluon distribution rises with 
$x$ due to rapid falloff of the distribution itself.
The prompt photon data were often used to better confine 
the gluon distribution at 
high $x$, but the prompt photon production data, 
which appeared recently, turned out to be in disagreement with  
the earlier data (see review \cite{Werlen:1999fn}).
Besides, it was shown, that in the theoretical analysis of this process
large uncertainties occur (see review \cite{Laenen:2000ii}). 
For these reasons one cannot use the prompt
photon data for pinning down the gluon distribution
in a consistent way.
In our analysis the gluon distribution at low $x$ is
determined by 
by the slope of the structure function $F_2$ on $Q$
(see Ref.\cite{Prytz:1993vr}) and at high $x$, 
from the partons momentum conservation. The experimental errors on the 
sea quarks distributions are also rather large, since we did not use 
in the analysis the Drell-Yan process data.

Unfortunately, the obtained PDFs and their errors suffers from definite
model dependence. For example, if one releases the constraint 
$a_{\rm su}=a_{\rm sd}=a_{\rm ss}$, the 
quarks distributions errors at low $x$ rise significantly.
Analogous effect is observed, if more polynomial factors 
are added to the starting PDFs. Such model dependence is inevitable, 
since it is impossible to determine a continuous distribution from 
limited number of measurements without additional constraints.
The model dependence is stronger for the PDFs
correlated with another PDFs, e.g., for the sea and valence quarks 
distributions, while the model dependence is weak for the sum of these 
distributions. The gluon distribution is also insensitive 
to the variation of the quark distributions due to rather
weak correlation with the latter (see Table~\ref{tab:pdfcor}).

The MRST and CTEQ PDFs are given in 
Fig.~\ref{fig:pdfs} in comparison with ours, although the comparison 
is incomplete, since the errors on the MRST and CTEQ PDFs 
are unknown. Note, that the difference between 
the MRST and CTEQ PDFs almost everywhere is smaller, than 
our PDFs errors. At high $x$ it may occur due to those 
collaborations use in the analysis more data, but more probable 
explanation is that the MRST and CTEQ collaborations get 
similar results due to they use similar data sets.
In particular, this means, that 
the difference between the MRST and CTEQ PDFs cannot be used 
as the estimate of the PDFs uncertainty. In the whole, with the account 
of our PDFs errors, there is no striking disagreement of our 
PDFs with the MRST and CTEQ ones. Our gluon distribution 
is slightly higher, than the MRST one at low $x$, but this 
disagreement is statistically insignificant. Excess of our 
sea quarks distributions over the MRST and CTEQ ones at low $x$
is statistically significant, but there are several reasons for it.
Firstly, both collaborations use massless scheme for the account of 
the heavy quarks contribution, that can lead to the overestimation of this 
contribution, and the corresponding underestimation of the 
light quarks contribution at low $x$. Secondly, the MRST and CTEQ 
collaborations use in the analysis the CCFR neutrino data of 
Ref.\cite{Seligman:1997mc}, which confine the sea quarks 
contribution and which were recently corrected by the authors
just at low $x$ (see Ref.\cite{Bodek:2000yr}).
Finally, the discrepancy between the MRST and CTEQ 
PDFs is of the order of discrepancy between those PDFs and ours, 
i.e. one needs to perform a detailed analysis of all 
parametrizations to clarify this discrepancy.
Excess of the $u$- and $d$-quarks distributions over 
the MRST and CTEQ ones at $x\lesssim 0.3$ is most 
statistically significant. We checked, that that this excess 
occurs due the MRST and CTEQ 
collaborations renormalize the BCDMS data by 1-2\% downward.
Since we do not apply such renormalization, our parametrization 
for $F_2$, as well as the $u$- and $d$-quarks distributions, lays higher.
Besides, we applied the TMC correction and the correction on the 
Fermi-motion in deuterium, that also leads do the rise 
of the quarks distributions at moderate $x$.
Note, that this excess may help to explain the 
excess of the TEVATRON jet production cross section data 
at transverse energies of $E_{\rm T}=200-400~{\rm GeV}$
over the QCD predictions, since this 
cross section gets large contribution from the quark-quark
scattering at $x\sim 0.2$.

\begin{figure}
\centerline{\epsfig{file=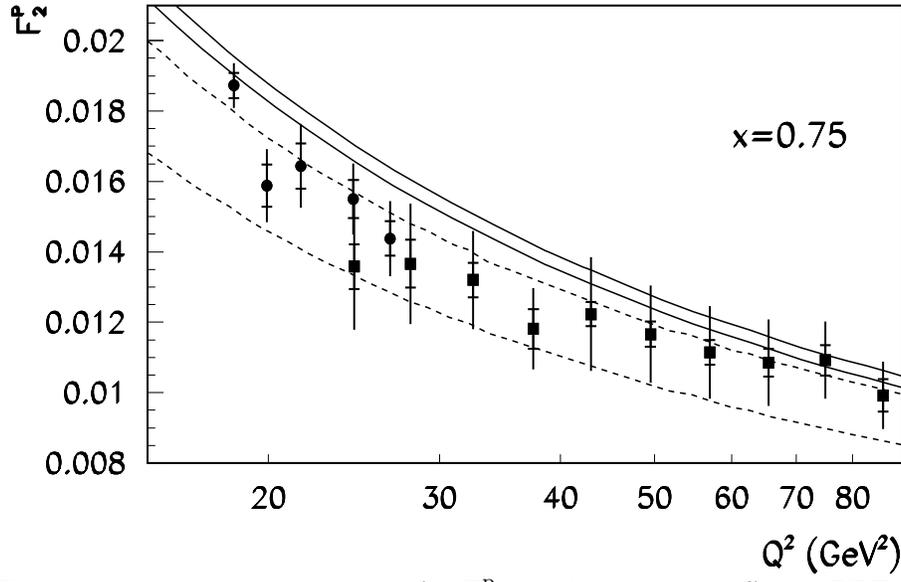,width=12cm,height=8cm}}
\caption{The $1\sigma$ experimental error bands for 
$F_2^{\rm p}$ calculated using different PDFs
(full line: our PDFs; dashes: the PDFs of 
Ref.\protect\cite{Botje:2000dj}).
Circles: the SLAC data; squares: the BCDMS ones.}
\label{fig:f2p}
\end{figure}

\begin{figure}
\centerline{\epsfig{file=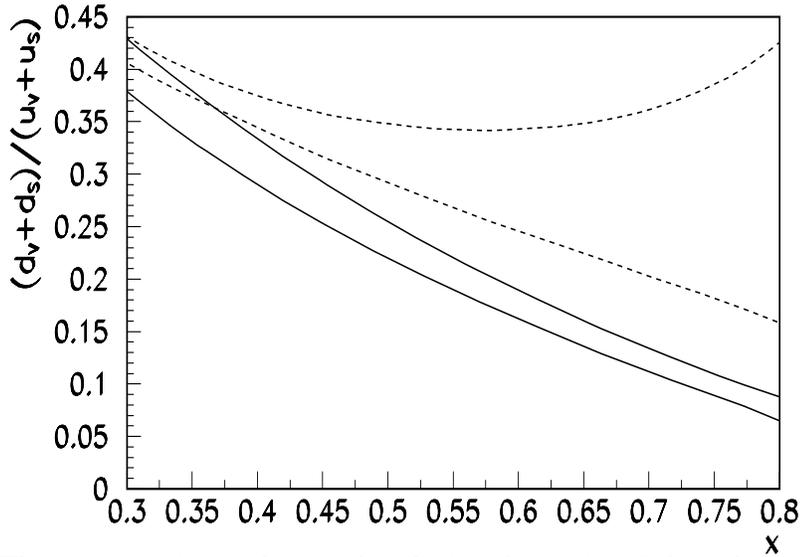,width=12cm,height=8cm}}
\caption{The $1\sigma$ experimental error bands for 
the ratio of $d$- and $u$-quarks distributions at 
$Q^2=9~{\rm GeV}^2$ (full lines: our PDFs; dashes:
the PDFs of Ref.\protect\cite{Botje:2000dj}.}
\label{fig:du}
\end{figure}

The comparison of our PDFs errors with the errors on PDFs 
of Ref.\cite{Botje:2000dj} is given in Fig.~\ref{fig:bot}.
One can see that, despite of that in the analysis of Ref.\cite{Botje:2000dj}
an additional NMC data on the 
neutron and proton structure functions ratio and the 
CCFR neutrino data are used, our PDFs errors are smaller generally. 
We ascribe this difference to that in the analysis
of Ref.\cite{Botje:2000dj} the SCE was used in the fit.
This conclusion is supported by the comparison of the 
structure function $F_2$ band, calculated from the  
PDFs of Ref.\cite{Botje:2000dj}, with the data used in that fit. 
The comparison is given in Fig.~\ref{fig:f2p}.
One can see that the most left point error  
is smaller, than the error on the $F_2$ parametrization 
of Ref.\cite{Botje:2000dj} for 
this point, i.e. SCE applied for that analysis
uses information given by this measurement inefficiently.
The qualitative explanation
of such behaviour of the SCE is that for this estimator 
the fitted parameters 
systematic errors are basically determined by the data points with 
the largest systematic errors. The CME used in our analysis 
is more efficient, than SCE and, as one can 
conclude from Fig.~\ref{fig:f2p}, our error on the $F_2$ parametrization
is basically confined by the point with the lowest 
systematic error. The difference of the SCE and CME PDFs errors 
is more the more  
is the relative contribution of the systematic errors to the total 
one. As a consequence this difference 
is especially large for the $u$-quark distribution and 
it is demonstrative, that the error on the $u$-quark distribution of
Ref.\cite{Botje:2000dj} almost coincide with the 
$u$-quark distribution errors obtained from our SCE fit
(see Fig.~\ref{fig:bot}). The error on $d$- and $u$-quarks 
distributions ratio at high $x$ given by our PDFs  
is also smaller, as compared with this error given by the 
PDFs of Ref.\cite{Botje:2000dj} (see Fig.~\ref{fig:du}).

\section{The theoretical uncertainties}
\label{sec:theor-uns}

The theoretical uncertainties inherent for a phenomenological analysis
cannot be ultimately defined, since in the study progress 
the set of such uncertainties may increase or
decrease. In our analysis we accounted for the following 
sources of the theoretical uncertainties:
\begin{itemize}
\item[{ ${\rm MC}$}] -- the change of the $c$-quark mass by 0.25 GeV;

\item[{ ${\rm SS}$}] -- the change of the strange sea suppression factor 
by 0.1, in line with the estimate given by 
the ${\rm NuTeV}$ collaboration \cite{Adams:1999sx};

\item[{ ${\rm TS}$}] -- the change of the heavy quarks threshold values
from $m_{\rm c,b}$ to $\sqrt{6.5}m_{\rm c,b}$, in accordance with the 
consideration of Sec.\ref{sec:theory};

\item[{ ${\rm RS}$}] -- the change of the renormalization scale 
in evolution the equations from $Q/2$ to $2Q$;

\item[{ ${\rm DC}$}] -- the change of the deuterium nuclear model 
based on the account of Fermi-motion \cite{Atwood:1973zp}
on the phenomenological model of Ref.\cite{Gomez:1994ri}. 
In view of the discussion of Refs.\cite{Melnitchouk:1999un,Yang:2000ew}
on the applicability of the model of Ref.\cite{Gomez:1994ri} 
to the light nuclei, 
one may suppose that this change leads to the overestimation of the 
corresponding error.
\end{itemize}
These changes were made in turn and the fitted parameters shifts
for each change were taken as the theoretical errors on the parameters. 
Sometimes in other similar analysis the PDFs theoretical errors due to 
the $\alpha_{\rm s}$ and HT uncertainties are estimated using the same approach. 
In our analysis these errors are included in the total experimental errors, since 
both $\alpha_{\rm s}$ and the HT contribution are fitted.
We underline, that the scales of the considered theoretical errors are rather 
conventional, 
since they are based on the ``reasonable'' estimates of the model uncertainties.
For this reason the theoretical errors should be accounted for with certain cautions.

\section{The \boldmath{$\alpha_{\rm s}$} value and the HT contribution}
\label{sec:alphag}

We obtained from the fit the value 
$\alpha_{\rm s}(M_{\rm Z})=0.1165\pm0.0017({\rm stat+syst})$.
The experimental error on $\alpha_{\rm s}$
obtained in our analysis is two times less, than in the NLO
analysis of similar data set described in Ref.\cite{Santiago:1999pr},
where the value 
$\alpha_{\rm s}(M_{\rm Z})=0.1160\pm0.0034({\rm exp})$ was obtained.
The contributions of separate sources of the theoretical errors
on our value of $\alpha_{\rm s}(M_{\rm Z})$ are given in 
Table~IV.
One can see, that the largest 
contributions give uncertainties of the QCD renormalization scale 
and the heavy quarks threshold values (especially for $b$-quark).
Combining all these contributions in quadrature, we obtain
\begin{equation}  
\alpha_{\rm s}(M_{\rm Z})=0.1165\pm0.0017({\rm stat+syst})
\pm^{0.0026}_{0.0034}({\rm theor}),
\label{eqn:alpha-glob}
\end{equation}  
which agrees with the modern world average
$\alpha_{\rm s}(M_{\rm Z})=0.1184\pm0.0031$
given in Ref.\cite{Bethke:2000ai}. 
Our estimate of the $\alpha_{\rm s}$ value is 
insensitive to the complication of the PDFs form, since 
it is almost uncorrelated with the PDFs parameters, in particular, 
with the gluon distribution ones (see Table~\ref{tab:pdfcor}).

As it was recently reported in Ref.\cite{Vogt:1999ik}, 
the net partons momentum for the PDFs, obtained from the data set 
similar to one used in our analysis, is not equal to 1, if 
one does not cut the data with $Q^2\lesssim 10~{\rm GeV}^2$.
In particular, the net partons momentum obtained from  
the analysis of the world charged leptons DIS 
data with $Q^2\ge 3~${\rm GeV}$^2$ is 
$<x>\approx1.08\pm0.02$, as it is given in Ref.\cite{Vogt:1999ik}.
The conclusion drawn from this observation is 
that the DIS data at low $Q$ are irrelevant 
for the NLO QCD analysis and reliable results can be 
obtained from the fit to the data with  
$Q^2\ge 10~{\rm GeV}^2$, $W^2\ge 10~{\rm GeV}^2$ only. 
The value of $\alpha_{\rm s}(M_{\rm Z})=0.114\pm0.002$
obtained in this analysis differs from ours.
In order to perform comparison with this result, 
we repeated our fit without imposing the momentum 
conservation constraint on the PDFs and obtained that 
at $Q^2=9~{\rm GeV}^2$ the net partons momentum is   
$<x>=0.979\pm0.029$, which agrees with 1 and differs 
from the results of Ref.\cite{Vogt:1999ik}.
For this reason we
cannot support the conclusion of Ref.\cite{Vogt:1999ik} 
about irrelevance of the low $Q$ charged leptons DIS data
for the NLO QCD analysis.
For more detailed comparison we performed the
test fit with the cuts of Ref.\cite{Vogt:1999ik}
and also obtained the lower value
$\alpha_{\rm s}(M_{\rm Z})=0.1098\pm0.0055$,
but with the error, which is significantly larger, than 
one obtained in Ref.\cite{Vogt:1999ik}, and 
the $\alpha_{\rm s}$ value obtained in this test 
fit is in agreement with (\ref{eqn:alpha-glob})
within the errors. 
The observed difference of the $\alpha_{\rm s}$ errors
evidently occurs due to in  
our analysis we simultaneously fit both the $\alpha_{\rm s}$
value and the HT contribution to $F_2$. As it was shown in 
Refs.~\cite{Alekhin:1999hy,Alekhin:2000iq}, the latter are  
strongly correlated, that certainly leads to the rise of 
the parameters errors. In support of this conclusion, 
if in our test fit the HT contribution is fixed, 
the $\alpha_{\rm s}$ error
falls from 0.0055 to 0.0014. However, the results of the  
fit with the HT fixed are model dependent and essentially the 
decrease of the experimental error is accompanied by the 
uncontrolled rise of the theoretical errors. 

\begin{figure}
\centerline{\epsfig{file=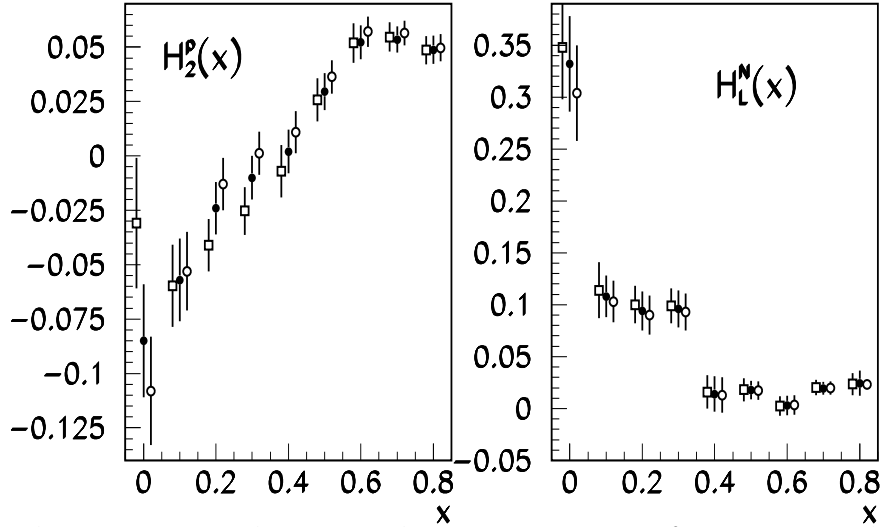,width=12cm,height=7cm}}
\caption{The twist 4 contribution to the 
proton structure function $F_2$ and to the nucleon 
structure function $F_{\rm L}$ (full circles: 
$\mu_{\rm R}=Q$; open circles: $\mu_{\rm R}=2Q$; 
squares: $\mu_{\rm R}=Q/2$). For better view the points 
corresponding to different $\mu_{\rm R}$
are shifted to left-right along the $x$-axis.}
\label{fig:hts}
\end{figure}

The HT contributions to the nucleon structure functions 
$F_{\rm L}$ and to the proton and neutron structure functions
$F_2$ are given in Fig.~\ref{fig:hts} and in Table~\ref{tab:hts}. 
It is interesting that up to minimal $x$ the twist 4 contribution 
to the structure function $F_2$ is non-zero,
that coincides with the results of 
Ref.\cite{Arneodo:1993kz} on the analysis of the NMC 
data. The deviation of the $F_{\rm L}$ twist 4 contribution 
off zero at low $x$ is even more significant.
As one can see from Table~\ref{tab:hts}, the HT contributions 
to $F_2$ and to $F_{\rm L}$ at low $x$ are very 
sensitive to the approach used to account for the systematic errors
on data. This is due to at low $x$ the HT contributions are 
determined from the comparison of the data at the 
kinematical edges of different experiments, 
where the systematic errors are largest as a rule. 
Note, that the HT parameters errors obtained in
the CME fit are 2-3 times smaller, than in the SCE fit, 
as well as the PDFs parameters errors.

The twist 4 contributions obtained at the different values of the QCD 
evolution equations renormalization scale $\mu_{\rm R}$ 
are given in Fig.~\ref{fig:hts}. The evident dependence of 
$H_2$ on $\mu_{\rm R}$ at low $x$ indicates that in this 
$x$-region the twist 4 contribution to $F_2$ can simulate the 
effect of the NNLO corrections to the splitting functions
$P$. Analogous effect for the structure function $xF_3$
was demonstrated in Ref.~\cite{Alekhin:2000af}, while the direct 
observation of the re-tuning of the twist 4 contribution to $xF_3$
due to the account of the NNLO corrections was reported in 
Ref.~\cite{Kataev:1998nc}. At the same time the $\mu_{\rm R}$
dependence of $H_{\rm L}$ and of $H_2$ at high $x$ 
is not so strong. The explanation of such behaviour 
is given in Ref.~\cite{Alekhin:1999kt}. As it was also shown there,
due to the HT contribution can partially absorb the NNLO corrections
effects, the $\mu_{\rm R}$ dependence of the $\alpha_{\rm s}$
value obtained in the simultaneous fit of the HT contribution
and $\alpha_{\rm s}$
is weaker, than in the fit with the HT contribution fixed. In particular, 
due to this absorption, the $\alpha_{\rm s}$ renormalization scale error 
obtained in our analysis is smaller, than in the analysis of 
Ref.~\cite{Vogt:1999ik}.
 
\begin{figure}
\centerline{\epsfig{file=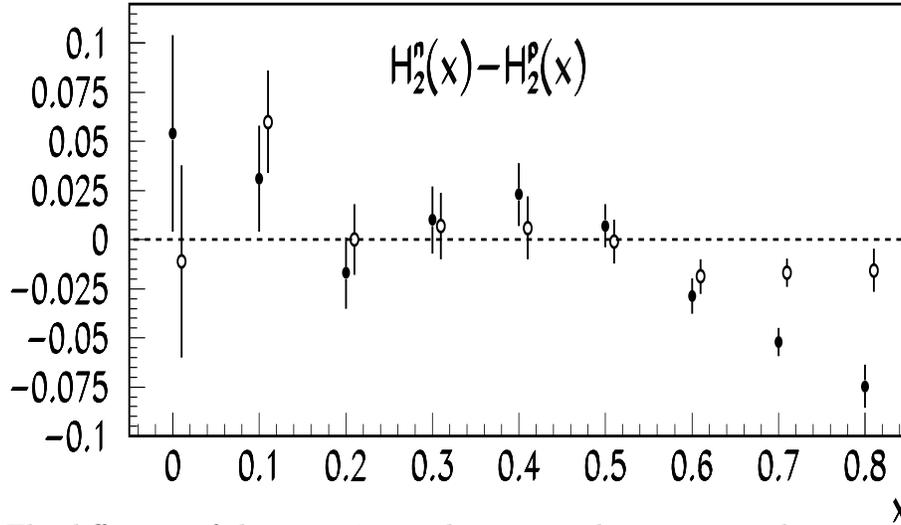,width=12cm,height=7cm}}
\caption{The difference of the twist 4 contributions to the 
neutron and proton structure functions $F_2$ obtained in the 
fits using the different deuterium models (full circles: the 
Fermi-motion model; open circles: the model of 
Ref.\protect\cite{Gomez:1994ri}.
For better view the points corresponding to different models 
are shifted to left-right along the $x$-axis.}
\label{fig:ht-n}
\end{figure}

The difference of the HT contributions to the proton and neutron 
structure functions $F_2$ is given in Fig.~\ref{fig:ht-n}.
One can see, that at low $x$ these contributions coincide
within errors. This is in disagreement with the results of 
Ref.~\cite{Szczurek:2000wp}.In that paper the data on the 
difference of the proton and neutron structure functions $F_2$ 
are compared with the calculations based on the standard PDFs
and found to be lower than that calculations at 
$x\sim 0.3$. This discrepancy was attributed to 
the existence of the large HT 
contribution to the difference of the proton and neutron 
structure functions $F_2$. 
We do observe the statistically significant deviation of 
$H_2^{\rm n}-H_2^{\rm p}$ off zero, but at $x\sim 0.7$ instead of 
$x\sim 0.3$. Unfortunately, this difference strongly depends on the 
deuterium nuclear corrections model at large $x$ 
(see Fig.~\ref{fig:ht-n}) and in order to 
obtain a reliable estimate of the twist 4 contribution to $F_2^{\rm n}$ 
an additional comparative analysis of the deuterium models is needed.

\section{The parton luminosities at the FNAL and LHC colliders}
\label{sec:lum}

\begin{figure}
\centerline{\epsfig{file=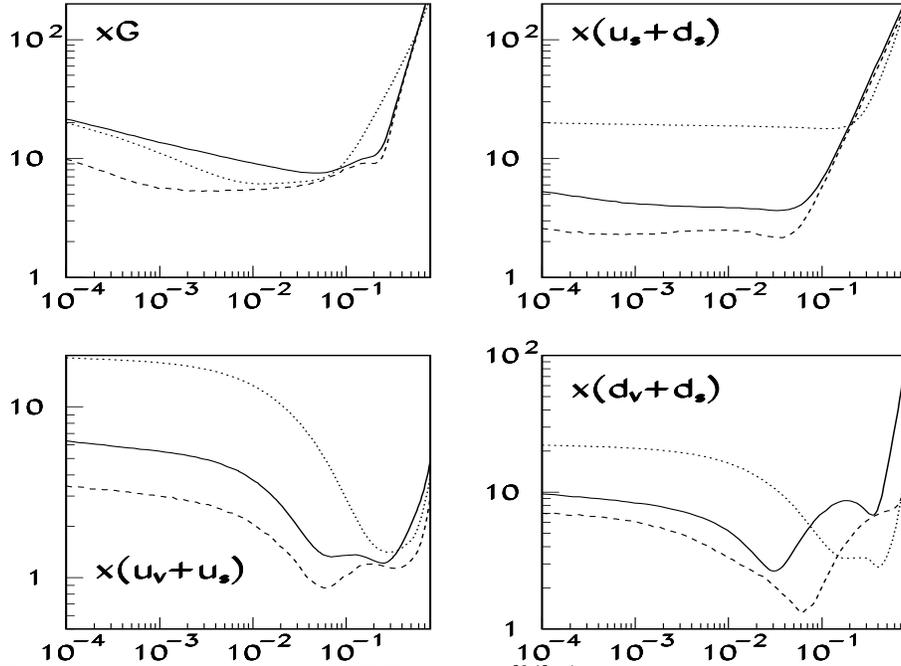,width=12cm,height=9cm}}
\caption{The relative experimental PDFs errors [\%] (full lines: the total errors, 
dashes: the experimental ones). For comparison the relative 
experimental errors on the PDFs of Ref.\protect\cite{Alekhin:1999za} are also 
given (dots).}
\label{fig:pdfs_err}
\end{figure}

\begin{figure}
\centerline{\epsfig{file=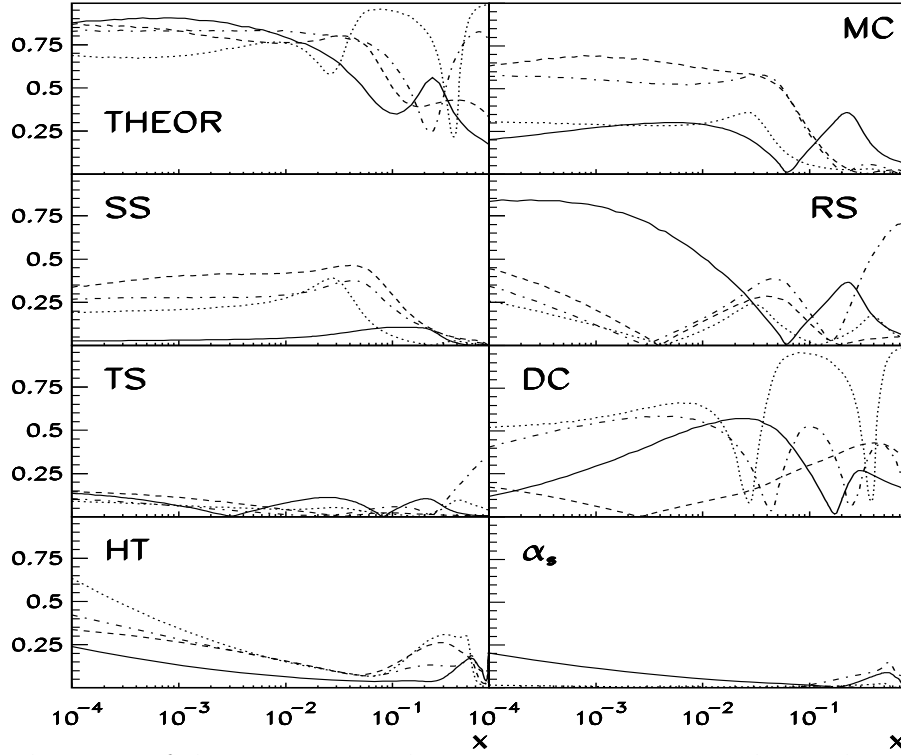,width=12cm,height=10cm}}
\caption{The ratios of the PDFs errors due to separate sources 
to the total PDFs errors (full lines: the gluon distribution;
dashes: the non-strange sea one; dots: the $d$-quark one; 
dotted-dashes: the $u$-quark one).
THEOR means the sum of the MC,SS,RS,TS, and DC 
contributions.}
\label{fig:pdfs_stat}
\end{figure}

All errors on the hard processes cross sections due to the PDFs uncertainties 
are concentrated in the parton luminosities (PLs), defined as 
$$
L_{ij}(M)=\frac{1}{s}\int^1_{\tau}\frac{dx}{x} q_i(x,M^2) 
q_j(\tau/x,M^2),  
$$
where $s$ is the s.c.m. energy squared; $M$ is the produced mass; 
$\tau=M^2/s$; $i$ and $j$ mark the parton species.
Since the PLs errors strongly depend
on the latter, one is to estimate the impact of the PDFs errors on the 
calculated cross sections errors in each particular case.
Our PDFs total errors, comprised of the theoretical errors combined 
with the experimental ones are given in Fig.~\ref{fig:pdfs_err}.
Despite of that the data set used for the extraction of our PDFs is limited
by the DIS data, the PDFs errors are rather small at low $x$, i.e. in the  
region especially important for the FNAL and LHC experiments.
The valence quarks distributions errors are small at high $x$ also
(see Fig.~\ref{fig:pdfs_err}). The experimental errors dominate
for the sea and gluon distributions at high $x$ only (note, that 
this is not the case for the SCE fit, as one can see from the comparison 
of Fig.~\ref{fig:pdfs_err} and Fig.~\ref{fig:bot}).
As one can see from Fig.~\ref{fig:pdfs_stat},
the dominating source for the gluon distribution at low $x$ 
is the RS uncertainty, for the sea distribution at low $x$, 
the MC one, for the $d$-quark distribution, the DC one. 
Remind, that in our analysis the errors due to the uncertainties 
of the $\alpha_{\rm s}$
value and the HT contribution are included into the experimental error.
To estimate their contribution to the total error we re-calculated the 
PDFs dispersions fixing the $\alpha_{\rm s}$ value and the HT
contribution by turn, then extracted obtained dispersions from the 
nominal dispersions 
calculated with these parameters released. The square roots of these 
differences were taken as the PDFs errors due to the $\alpha_{\rm s}$ and 
the HT uncertainties respectively. The ratios of these errors to the total 
PDFs errors are also given in Fig.~\ref{fig:pdfs_stat}. One can see, 
that the $\alpha_{\rm s}$ uncertainty affects the gluon distribution 
only, while the HT uncertainty contributes to the errors of all PDFs.

\begin{figure}
\centerline{\epsfig{file=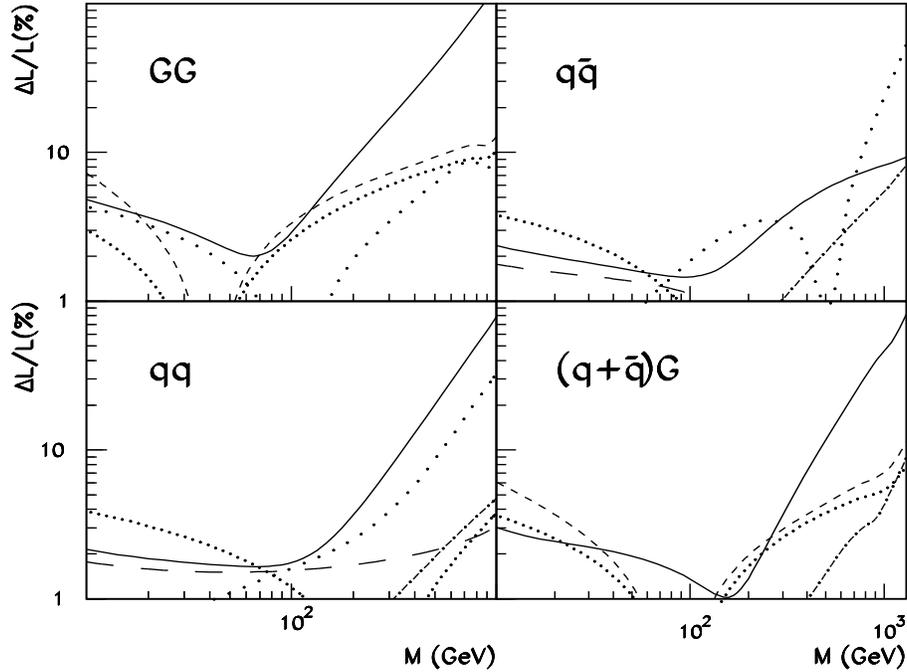,width=12cm,height=9cm}}
\caption{The relative errors on selected PLs for the FNAL 
collider (full lines: experimental errors, short dashes:
RS; dotted-dashes: TS; sparse dots: DC; dense dots: MC;
long dashes: SS). Other notations:
$L_{\rm qq}=L_{\rm uu}+L_{\rm dd}+L_{\rm du}$; 
$L_{\rm q\bar q}=L_{\rm u \bar d}+L_{\rm d \bar u}$; 
$L_{\rm (q+\bar q)G}=L_{\rm uG}+L_{\rm \bar uG}+L_{\rm dG}+L_{\rm \bar dG}$.}
\label{fig:lums-fnal}
\end{figure}

\begin{figure}
\centerline{\epsfig{file=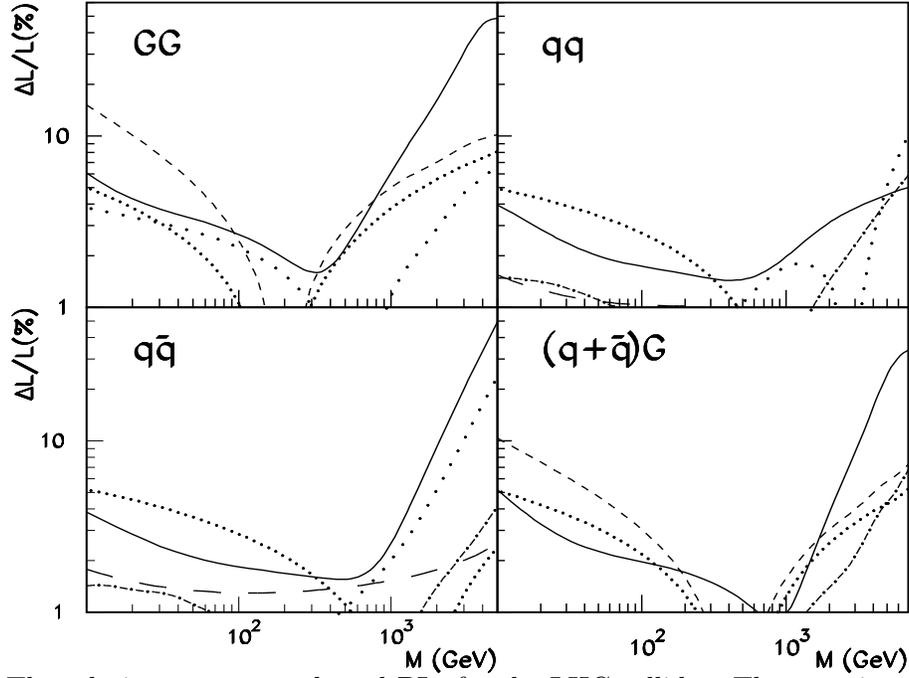,width=12cm,height=9cm}}
\caption{The relative errors on selected PLs for the LHC
collider. The notations are the same as in Fig.~\protect\ref{fig:lums-fnal}.} 
\label{fig:lums-lhc}
\end{figure}

The errors on the PLs relevant for the most common processes 
at the energy of the FNAL collider are given in Fig.~\ref{fig:lums-fnal}. 
The upper limit of the pictures was chosen 
so that the PLs at the upper limit is 
$\sim 0.01$~1/pb, i.e. corresponds to the maximal sensitivity of 
the planned experiments. One can see, that at the FNAL collider energy
the theoretical errors dominate over the experimental ones at  
$M\lesssim 0.2~{\rm TeV}$ and vice versa at 
$M\gtrsim 0.2~{\rm TeV}$. The total PLs errors for the FNAL collider
generally do not exceed 10\% at $M\lesssim 0.2~{\rm TeV}$, while
for the quark-antiquark PL the total error is smaller, than 
10\% almost for all $M$ in question. The PLs pictures for the LHC energy, 
given in Fig.~\ref{fig:lums-lhc}, approximately reproduce the FNAL 
pictures with the produced mass $M$ scaled in 5 times and the quark-antiquark
PL replaced by the quark-quark PL.

Due to the PDFs correlations generally are not small 
(see Fig.~\ref{fig:pdfs_corr}), 
the account of these correlations may affect the 
calculated hard processes cross sections errors. 
In some cases the PLs errors 
may cancel in their ratio, as in the example given 
in Table~\ref{tab:wz}.
Calculating the theoretical errors on the hard processes cross sections
one is also to take into account the correlations of PDFs with the 
elementary processes cross sections, if the latter depend on the parameters 
responsible for the PDFs theoretical uncertainties. Besides,  
the RS PDFs uncertainty may be compensated by the NNLO 
corrections to the elementary processes cross sections.

\section{Conclusion}

Significant part of the studies planned for the  
next generation hadron-hadron and lepton-hadron colliders is devoted to the 
precise Standard Model checks (see, e.g., review \cite{QCD}). 
Such studies certainly imply careful control of all possible uncertainties, 
including the PDFs errors.
The PDFs obtained in our analysis are supplied by the experimental 
and theoretical errors and can be used for the correct estimate of the 
calculated hard processes cross sections uncertainties, necessary for 
a precise phenomenological comparison aiming to detect a 
manifestation of new physics
(e.g., compositness in proton-proton and electron-proton collisions,
the partons recombination at low $x$, precise determination of the 
$W$ and $Z$ masses, etc.). A particular feature of our PDFs is that 
they were obtained using efficient estimator and, as a result, have 
minimal errors. The convenient code allowing to account 
for the PDFs uncertainties in the Monte Carlo calculations is 
accessible through the computer network\footnote{The WWW address is 
http://www.ihep.su/$\tilde{~}$alekhin/pdf99}.
Using the current version of this code one can obtain the  
random Gaussian smeared PDFs values with the account 
of the experimental and the theoretical uncertainties and their correlations.
The special parameters allow one to scale the dispersions corresponding to the 
separate sources of the PDFs uncertainties to give user the possibility 
to study effects of each uncertainty and vary the confidence level of
the errors on the calculations results.

Author is indebted to A.L.~Kataev and S.A.~Kulagin for the careful 
reading of the manuscript and valuable comments, 
S.~Keller and W.J.~Stirling for 
the interesting discussions. The work was partially supported by 
the RFBR grant 00-02-17432. The final part of the work was completed during 
the visit to CERN and author is grateful to the staff of the TH division for 
providing good working conditions.

\newpage 

\newpage 
\begin{table}
\caption{Total number of points (NDP), number of independent 
sources of systematic errors (NSE), $\chi^2/{\rm NDP}$ 
and the net residual $R$ for each experiment (standard deviation of $R$ is 
given in parenthesis). Also the renormalization factors
$\xi$ for the old SLAC experiments are given.}
\begin{center}
\scriptsize
\begin{tabular}{cccccc} 
&\multicolumn{2}{c}{${\rm NDP}/(1-\xi)$[\%]}&&& \\ \cline{2-3}
experiment&proton&deuterium&${\rm NSE}$&
$\chi^2/{\rm NDP}$&$R$\\  \hline
${\rm SLAC}$-${\rm E}$-${\rm 49A}$  &58~/~$1.8\pm1.3$ &58~/~-$0.4\pm1.2$   &3&0.52&--0.05(0.23)   \\ 
${\rm SLAC}$-${\rm E}$-${\rm 49B}$  &144~/~$2.0\pm1.3$&135~/~$-0.1\pm1.3$ &3&1.20&0.22(0.29)     \\  
${\rm SLAC}$-${\rm E}$-${\rm 87}$   &90~/~$2.0\pm1.2$&90~/~$0.2\pm1.2$   &3&0.91&0.01(0.37)   \\  
${\rm SLAC}$-${\rm E}$-${\rm 89A}$   &66~/~$4.2\pm1.8$&59~/~$1.2\pm1.9$  &3&1.34&--0.18(0.45)    \\  
${\rm SLAC}$-${\rm E}$-${\rm 89B}$  &79~/~$1.5\pm1.2$&62~/~$-0.7\pm1.2$   &3&0.82&0.46(0.49)   \\  
${\rm SLAC}$-${\rm E}$-${\rm 139}$   &--&16~/~$1.0\pm1.2$  &3&0.64&--0.10(0.43)    \\ 
${\rm SLAC}$-${\rm E}$-${\rm 140}$  &--&26   &4&0.89&0.51(0.86)   \\ 
${\rm BCDMS}$  &351&254 &9&1.15&0.07(0.68)        \\ 
${\rm NMC}$    &245&245 &13&1.32&0.05(0.62)   \\  
${\rm H1(94)}$    &--&147 &5&0.96&0.11(0.25)   \\ 
${\rm ZEUS(94)}$  &--&188 &20&2.14&0.32(0.34)   \\
${\rm FNAL}$-${\rm E}$-${\rm 665}$    &47&47 &10&1.23&0.38(0.38)   \\  
Total  &1080&1327 &79&1.20&0.12(0.22)     \\ 
\end{tabular}
\label{tab:global}
\end{center}
\normalsize
\end{table}

\begin{table}
\caption{The fitted $\alpha_{\rm s}$ and the 
PDFs parameters values (I: the CME fit; II: the SCE fit;
III: the fit with the statistical and systematic errors
combined in quadrature). Given errors on the parameters
include both statistical and systematic errors, pure 
statistical errors are given in parenthesis.}
\scriptsize
\begin{center}
\begin{tabular}{lcccc}   
       && I &II & III\\ \hline
Valence quarks:&&&&        \\
            &$a_{\rm u}$&$0.693\pm0.033(0.027)$&$0.715\pm0.114(0.029)$
&$0.703\pm0.035$        \\
       &$b_{\rm u}$&$3.945\pm0.050(0.039)$&$4.119\pm0.257(0.038)$
&$4.037\pm0.049$        \\
       &$\gamma_2^{\rm u}$&$1.29\pm0.44(0.37)$&$1.39\pm1.86(0.40)$
&$1.42\pm0.49$    \\
       &$a_{\rm d}$&$0.725\pm0.086(0.082)$&$0.703\pm0.172(0.094)$
&$0.717\pm0.13$        \\
       &$b_{\rm d}$&$4.93\pm0.13(0.12)$&$4.83\pm0.27(0.17)$
&$5.00\pm0.17$        \\
Gluon:&&&&        \\
&$a_G$&$-0.225\pm0.035(0.031)$&$-0.169\pm0.065(0.029)$
&$-0.135\pm0.044$        \\
       &$b_{\rm G}$&$6.1\pm2.1(1.8)$&$4.9\pm5.6(1.7)$
&$4.07\pm1.3$        \\
       &$\gamma_1^{\rm G}$&$-2.63\pm0.83(0.71)$&$-3.41\pm0.99(0.45)$
&$-4.06\pm0.48$    \\
       &$\gamma_2^{\rm G}$&$4.7\pm2.9(2.4)$&$4.44\pm3.4(1.3)$
&$5.41\pm1.2$    \\
Sea quarks:&&&&        \\
&$A_{\rm S}$&$0.166\pm0.011(0.0095)$ &$0.167\pm0.025(0.011)$
&$0.167\pm0.017$       \\
&$a_{\rm sd}$&$-0.1987\pm0.0067(0.0050)$&$-0.1853\pm0.0181(0.0050)$
&$-0.1833\pm0.0075$       \\
       &$b_{\rm sd}$&$5.1\pm1.4(1.3)$ &$5.4\pm2.8(1.4)$
&$4.9\pm2.1$       \\
       &$\eta_{\rm u}$&$1.13\pm0.11(0.087)$ &$1.10\pm0.23(0.086)$
&$1.16\pm0.16$       \\
       &$b_{\rm su}$&$10.29\pm0.97(0.81)$ &$10.56\pm3.2(0.83)$
&$11.2\pm1.1$       \\
&&&&        \\
       &$\alpha_{\rm s}(M_{\rm Z})$&$0.1165\pm0.0017(0.0014)$ 
&$0.1138\pm0.0044(0.0021)$&$0.1190\pm0.0036$ \\
\end{tabular}
\end{center}
\normalsize
\label{tab:pdfpars}
\end{table}

\begin{table}
\caption{The correlation coefficients for the starting PDFs parameters.
The largest coefficients are printed in bold.}
\tiny
\begin{tabular}{cccccccccccccccc}
&$a_{\rm u}$ & $b_{\rm u}$ & $\gamma_2^{\rm u}$ & $a_{\rm d}$ & $b_{\rm d}$ 
& $A_{\rm S}$ & $a_{\rm sd}$ & $b_{\rm sd}$ &
$\eta_{\rm u}$ & $b_{\rm su}$ & $a_{\rm G}$ & $b_{\rm G}$ & 
$\gamma_1^{\rm G}$ & $\gamma_2^{\rm G}$ & $\alpha_{\rm s}(M_{\rm Z})$  \\ 
$a_{\rm u}$   &1.00 &&&&&&&&&&&&&& \\
$b_{\rm u}$&--{\bf 0.84}&1.00 &&&&&&&&&&&&& \\
$\gamma_2^{\rm u}$
&\bf{-0.97}&\bf{0.92}&1.00 &&&&&&&&&&&& \\
 $a_{\rm d}$ 
&--0.09&--0.09&0.05&1.00 &&&&&&&&&&& \\
 $b_{\rm d}$ 
&--0.21&0.02&0.19&0.71&1.00 &&&&&&&&&& \\
 $A_{\rm S}$ 
&--0.14&0.34&0.24&\bf{--0.86}&--0.54&1.00 &&&&&&&&& \\
 $a_{\rm sd}$ 
&0.58&--0.45&--0.55&0.37&0.16&--0.46&1.00 &&&&&&&& \\
$b_{\rm sd}$ 
&--0.05&--0.10&0.00&\bf{0.97}&0.54&\bf{--0.88}&0.40&1.00 &&&&&&& \\
$\eta_{\rm u}$
&0.25&--0.13&--0.23&--0.69&--0.24&0.47&0.01&--0.78&1.00 &&&&&& \\
 $b_{\rm su}$ 
&\bf{0.83}&--0.74&\bf{--0.86}&--0.14&--0.16&--0.24&0.62&--0.10&0.44&1.00 &&&&& \\
$a_{\rm G}$
&0.23&--0.22&--0.23&0.37&0.20&--0.38&0.53&0.37&--0.21&0.18&1.00 &&&& \\
 $b_{\rm G}$ 
&0.18&--0.17&--0.20&0.11&0.17&--0.08&--0.10&0.06&--0.11&--0.02&0.27&1.00 &&& \\
 $\gamma_1^{\rm G}$ 
&--0.36&0.34&0.36&--0.45&--0.30&0.48&--0.52&--0.44&0.18&--0.30&\bf{--0.82}&--0.47&1.00 && \\
 $\gamma_2^{\rm G}$ 
&0.34&--0.34&--0.36&0.28&0.26&--0.32&0.15&0.23&--0.11&0.20&0.46&\bf{0.89}&--0.77&1.00&\\ 
 $\alpha_{\rm s}(M_{\rm Z})$
&0.22&--0.31&--0.18&0.01&--0.05&--0.05&0.04&--0.01&0.04&0.17&0.01&--0.39&0.03&--0.18&1.00\\
\end{tabular}
\label{tab:pdfcor}
\end{table}

\begin{table}
\begin{center}
\caption{The $\alpha_{\rm s}(M_{\rm Z})$ theoretical errors due to different sources.}
\begin{tabular}{cc} 
Source & Value \\ 
${\rm MC}$ & $\pm0.0003$ \\ 
${\rm SS}$ & $\pm0.0001$ \\ 
${\rm RS}$ & $\pm_{0.0024}^{0.0026}$ \\ 
${\rm TS}$ & $-0.0020$ \\ 
${\rm DC}$ & $-0.0012$ \\ 
\end{tabular}
\end{center}
\protect\label{tab:alpha-teor}
\end{table}

\begin{table}
\caption{The fitted twist 4 contributions 
(I: the CME fit; II: the SCE fit; III: the fit with 
the statistical and systematic errors combined in quadrature).
Given errors on the parameters
include both statistical and systematic errors, pure 
statistical errors are given in parenthesis.}
\scriptsize
\begin{center}
\begin{tabular}{ccccc}  
                &  $x$      &      I            & II                & III              \\  \hline
$H_2^{\rm p}$:&&   &   &   \\
&0.& $-0.085\pm0.026(0.020)$  & $-0.124\pm0.051(0.020)$  & $-0.132\pm0.035$  \\ 
&0.1          & $-0.057\pm0.019(0.014)$  & $-0.107\pm0.076(0.014)$  & $-0.094\pm0.021$ \\
&0.2          & $-0.024\pm0.012(0.0097)$  & $-0.057\pm0.049(0.010)$  & $-0.054\pm0.016$  \\ 
&0.3          & $-0.010\pm0.010(0.0089)$  & $-0.027\pm0.024(0.0090)$  & $-0.017\pm0.015$  \\ 
&0.4          & $0.002\pm0.010(0.0089)$  & $0.002\pm0.024(0.0090)$   & $-0.002\pm0.016$   \\ 
&0.5          & $0.0292\pm0.0085(0.0074)$   & $0.041\pm0.020(0.0079)$   & $0.025\pm0.015$   \\ 
&0.6          & $0.0522\pm0.0078(0.0069)$   & $0.068\pm0.017(0.0074)$   & $0.051\pm0.013$   \\ 
&0.7          & $0.0535\pm0.0061(0.0055)$   & $0.074\pm0.013(0.0058)$     & $0.056\pm0.010$     \\ 
&0.8          & $0.0488\pm0.0064(0.0061)$     & $0.0545\pm0.0085(0.0060)$     & $0.0471\pm0.0085$     \\  
$H_{\rm L}^{\rm N}$:&&   &   &   \\
&0.    & $0.332\pm0.046(0.033)$  & $0.13\pm0.11(0.033)$  & $0.028\pm0.061$  \\ 
&0.1    & $0.108\pm0.020(0.016)$  & $0.117\pm0.065(0.016)$  & $0.118\pm0.022$  \\ 
&0.2    & $0.094\pm0.019(0.015)$  & $0.145\pm0.047(0.015)$  & $0.097\pm0.021$  \\ 
&0.3    & $0.096\pm0.018(0.016)$  & $0.133\pm0.031(0.016)$  & $0.115\pm0.021$  \\ 
&0.4    & $0.014\pm0.017(0.015)$   & $0.040\pm0.027(0.015)$   & $0.033\pm0.019$   \\ 
&0.5    & $0.0179\pm0.0088(0.0068)$   & $0.023\pm0.014(0.0069)$   & $0.015\pm0.011$   \\ 
&0.6    & $0.0031\pm0.0094(0.0076)$   & $-0.016\pm0.024(0.0076)$   & $-0.0033\pm0.0089$   \\ 
&0.7    & $0.0195\pm0.0064(0.0056)$   & $0.008\pm0.016(0.0055)$     & $0.0134\pm0.0067$     \\ 
&0.8    & $0.024\pm0.012(0.012)$     & $0.01\pm0.023(0.012)$     & $0.012\pm0.014$     \\
$H_2^{\rm n}-H_2^{\rm p}$:&&   &   &   \\
&0.    & $0.054\pm0.050(0.041)$  & $0.045\pm0.112(0.041)$  & $0.095\pm0.077$  \\ 
&0.1    & $0.031\pm0.027(0.026)$  & $0.041\pm0.047(0.026)$  & $0.003\pm0.037$  \\ 
&0.2    & $-0.017\pm0.018(0.017)$  & $0.024\pm0.046(0.017)$  & $-0.014\pm0.024$  \\ 
&0.3    & $0.010\pm0.017(0.016)$  & $0.052\pm0.038(0.016)$  & $0.014\pm0.021$  \\ 
&0.4    & $0.023\pm0.016(0.015)$  & $0.047\pm0.037(0.015)$  & $0.036\pm0.019$  \\ 
&0.5    & $0.0068\pm0.011(0.010)$  & $0.009\pm0.026(0.011)$  & $0.019\pm0.016$  \\ 
&0.6    & $-0.029\pm0.0091(0.0086)$  & $-0.037\pm0.016(0.0092)$  & $-0.022\pm0.015$  \\
&0.7    & $-0.052\pm0.0073(0.0068)$  & $-0.073\pm0.014(0.0071$)  & $-0.055\pm0.011$  \\ 
&0.8    & $-0.075\pm0.011(0.010)$  & $-0.079\pm0.014(0.010)$  & $-0.073\pm0.013$  \\
\end{tabular}
\end{center}
\normalsize
\label{tab:hts}
\end{table}

\begin{table}
\caption{The relative errors on the PLs involved in the 
calculations of the $W$ and $Z$ production cross sections 
at the FNAL collider ($L_{\rm W}=L_{\rm u\bar d}+L_{\rm d\bar u}$, 
$L_{\rm Z}=L_{\rm u\bar u}+L_{\rm d\bar d}$,
$L_{\rm W/Z}=(L_{\rm u\bar d}+L_{\rm d\bar u})
/(L_{\rm u\bar u}+L_{\rm d\bar d})$).}
\begin{center}
\begin{tabular}{ccccccc}  
& stat+syst & ${\rm RS}$ & ${\rm TS}$ & 
${\rm SS}$ & ${\rm MC}$ & ${\rm DC}$ \\
$\Delta L_W (\%)$       &1.5 & -- & -- & 1.2 & 1.1  & 1.5\\
$\Delta L_{\rm Z} (\%)$ &1.2 & -- & -- & 1.2 & 1.1  & 1.5\\
$\Delta L_{W/Z} (\%)$ & 0.7 & -- & -- & -- & -- & --\\
\end{tabular}
\end{center}
\label{tab:wz}
\end{table}


\begin{thebibliography}{99}

\bibitem{Whitlow:1992uw}
L.~W.~Whitlow, E.~M.~Riordan, S.~Dasu, S.~Rock and A.~Bodek,
Phys.\ Lett.\  {\bf B282}, 475 (1992).

\bibitem{Benvenuti:1989rh}
A.~C.~Benvenuti {\it et al.}  [BCDMS Collaboration],
Phys.\ Lett.\  {\bf B223} (1989) 485;\\
A.~C.~Benvenuti {\it et al.}  [BCDMS Collaboration],
Phys.\ Lett.\  {\bf B237} (1990) 592.

\bibitem{Arneodo:1997qe}
M.~Arneodo {\it et al.}  [New Muon Collaboration],
Nucl.\ Phys.\  {\bf B483} (1997) 3
[hep-ph/9610231].

\bibitem{Adams:1996gu}
M.~R.~Adams {\it et al.}  [E665 Collaboration],
Phys.\ Rev.\  {\bf D54} (1996) 3006.

\bibitem{Aid:1996au}
S.~Aid {\it et al.}  [H1 Collaboration],
Nucl.\ Phys.\  {\bf B470}, 3 (1996)
[hep-ex/9603004].

\bibitem{Derrick:1996hn}
M.~Derrick {\it et al.}  [ZEUS Collaboration],
Z.\ Phys.\  {\bf C72}, 399 (1996)
[hep-ex/9607002].

\bibitem{Martin:2000ww}
A.~D.~Martin, R.~G.~Roberts, W.~J.~Stirling and R.~S.~Thorne,
Eur.\ Phys.\ J.\  {\bf C14}, 133 (2000)
[hep-ph/9907231].

\bibitem{Lai:2000wy}
H.~L.~Lai {\it et al.}  [CTEQ Collaboration],
Eur.\ Phys.\ J.\  {\bf C12}, 375 (2000)
[hep-ph/9903282].

\bibitem{Alekhin:1999za}
S.~I.~Alekhin,
Eur.\ Phys.\ J.\  {\bf C10} (1999) 395
[hep-ph/9611213].

\bibitem{Botje:2000dj}
M.~Botje,
Eur.\ Phys.\ J.\  {\bf C14}, 285 (2000)
[hep-ph/9912439].

\bibitem{Alekhin:2000es}
S.~I.~Alekhin, Preprint IFVE-2000-17 (2000) [hep-ex/0005042].

\bibitem{Altarelli:1982ax} 
B.~L.~Ioffe, V.~A.~Khoze and L.~N.~Lipatov,
``Hard Processes. Vol. 1: Phenomenology, Quark Parton Model,''
{\it  Amsterdam, Netherlands: North-Holland ( 1984) 340p}.

\bibitem{Wilson:1969zs}
K.~G.~Wilson,
Phys.\ Rev.\  {\bf 179}, 1499 (1969).

\bibitem{Georgi:1976ve}
H.~Georgi and H.~D.~Politzer,
Phys.\ Rev.\  {\bf D14}, 1829 (1976).

\bibitem{Bukhvostov:1983te}
A.~P.~Bukhvostov, E.~A.~Kuraev and L.~N.~Lipatov,
Yad.\ Fiz.\  {\bf 38}, 439 (1983).

\bibitem{Alekhin:2000iq}
S.~I.~Alekhin,
Eur.\ Phys.\ J.\  {\bf C12}, 587 (2000)
[hep-ph/9902241].

\bibitem{Gribov:1972ri}
V.~N.~Gribov and L.~N.~Lipatov,
Sov.\ J.\ Nucl.\ Phys.\ {\bf 15}, 438 (1972);\\
V.~N.~Gribov and L.~N.~Lipatov,
Sov.\ J.\ Nucl.\ Phys.\ {\bf 15}, 675 (1972);\\
G.~Altarelli and G.~Parisi, Nucl.\ Phys.\  {\bf B126}, 298 (1977);\\
Y.~L.~Dokshitzer, Sov.\ Phys.\ JETP {\bf 46}, 641 (1977).

\bibitem{SanchezGuillen:1991iq}
J.~Sanchez Guillen, J.~Miramontes, M.~Miramontes, G.~Parente and O.~A.~Sampayo,
Nucl.\ Phys.\  {\bf B353}, 337 (1991);\\
W.~L.~van Neerven and E.~B.~Zijlstra,
Phys.\ Lett.\  {\bf B272}, 127 (1991), ibid. {\bf B273}, 476 (1991),
ibid. {\bf B297}, 377 (1992);\\
W.~L.~van Neerven and E.~B.~Zijlstra,
Nucl.\ Phys.\  {\bf B382}, 11 (1992).

\bibitem{Larin:1997wd}
S.~A.~Larin, T.~van Ritbergen and J.~A.~Vermaseren,
Nucl.\ Phys.\  {\bf B427}, 41 (1994);\\
S.~A.~Larin, P.~Nogueira, T.~van Ritbergen and J.~A.~Vermaseren,
Nucl.\ Phys.\  {\bf B492}, 338 (1997)
[hep-ph/9605317];\\
A.~Retey and J.~A.~Vermaseren, hep-ph/0007294 (2000).

\bibitem{vanNeerven:2000ca}
W.~L.~van Neerven and A.~Vogt,
Nucl.\ Phys.\  {\bf B568}, 263 (2000)
[hep-ph/9907472];\\
W.~L.~van Neerven and A.~Vogt,
Phys.\ Lett.\  {\bf B490}, 111 (2000)
[hep-ph/0007362].

\bibitem{Kataev:1998nc}
A.~L.~Kataev, A.~V.~Kotikov, G.~Parente and A.~V.~Sidorov,
Phys.\ Lett.\  {\bf B417}, 374 (1998)
[hep-ph/9706534].

\bibitem{Kataev:1998ce}
A.~L.~Kataev, G.~Parente and A.~V.~Sidorov,
Nucl.\ Phys.\  {\bf B573}, 405 (2000)
[hep-ph/9905310].

\bibitem{Santiago:1999pr}
J.~Santiago and F.~J.~Yndurain,
Nucl.\ Phys.\  {\bf B563}, 45 (1999)
[hep-ph/9904344].

\bibitem{Vogt:1999ik}
A.~Vogt,
Nucl.\ Phys.\ Proc.\ Suppl.\  {\bf 79}, 102 (1999)
[hep-ph/9906337]

\bibitem{Furmanski:1982cw}
W.~Furmanski and R.~Petronzio,
Z.\ Phys.\  {\bf C11}, 293 (1982);\\
W.~Furmanski and R.~Petronzio,
Phys.\ Lett.\  {\bf B97}, 437 (1980);\\
G.~Curci, W.~Furmanski and R.~Petronzio,
Nucl.\ Phys.\  {\bf B175}, 27 (1980).

\bibitem{Alekhin:1999hy}
S.~I.~Alekhin,
Phys.\ Rev.\  {\bf D59}, 114016 (1999)
[hep-ph/9809544].

\bibitem{Bernreuther:1982sg}
W.~Bernreuther and W.~Wetzel,
Nucl.\ Phys.\  {\bf B197}, 228 (1982), 
ibid.{\bf B513}, 758 (1998) (Err);\\
S.~A.~Larin, T.~van Ritbergen and J.~A.~Vermaseren,
Nucl.\ Phys.\  {\bf B438}, 278 (1995)
[hep-ph/9411260];\\
K.~G.~Chetyrkin, B.~A.~Kniehl and M.~Steinhauser,
Phys.\ Rev.\ Lett.\  {\bf 79}, 2184 (1997)
[hep-ph/9706430].

\bibitem{Blumlein:1999sh}
J.~Blumlein and W.~L.~van Neerven,
Phys.\ Lett.\  {\bf B450} (1999) 417
[hep-ph/9811351].

\bibitem{Barnett:1996hr}
R.~M.~Barnett {\it et al.},
Phys.\ Rev.\  {\bf D54}, 1 (1996).

\bibitem{Martin:1991jd}
A.~D.~Martin, W.~J.~Stirling and R.~G.~Roberts,
Phys.\ Lett.\  {\bf B266}, 173 (1991).

\bibitem{INDUR} Yndurain, F.J.,
   ``Quantum Chromodynamics: an Introduction to the Theory of Quarks
and Gluons'', Springer-Verlag, 1983. 227p.

\bibitem{Matveev:1973ra}
V.~A.~Matveev, R.~M.~Muradian and A.~N.~Tavkhelidze,
Lett.\ Nuovo Cim.\  {\bf 7}, 719 (1973).

\bibitem{Brodsky:1973kr}
S.~J.~Brodsky and G.~R.~Farrar,
Phys.\ Rev.\ Lett.\  {\bf 31}, 1153 (1973).

\bibitem{Alekhin:1999kt}
S.~I.~Alekhin, 
Phys.\ Lett.\  {\bf B488}, 187 (2000)
[hep-ph/9912484].

\bibitem{Blumlein:1996rp}
J.~Blumlein, S.~Riemersma, M.~Botje, C.~Pascaud, F.~Zomer, W.~L.~van Neerven 
and A.~Vogt, in ``Hamburg 1995/1996, Future physics at HERA'', p.23 (1996)
[hep-ph/9609400].

\bibitem{MATHBOOK} ``Handbook of Mathematical Functions with Formulas, Graphs,
and Mathematical Tables'', ed. by M.~Abramowitz and I.~A.~Stegun., 
Dover, 1972. 1046p.  (Applied Math Series, 55)

\bibitem{Collins:1986mp}
J.~C.~Collins and W.~Tung,
Nucl.\ Phys.\  {\bf B278}, 934 (1986).

\bibitem{Witten:1976bh}
E.~Witten,
Nucl.\ Phys.\  {\bf B104}, 445 (1976).

\bibitem{Shifman:1978yb}
M.~A.~Shifman, A.~I.~Vainshtein and V.~I.~Zakharov,
Nucl.\ Phys.\  {\bf B136}, 157 (1978).

\bibitem{Gluck:1994dp}
M.~Gluck, E.~Reya and M.~Stratmann,
Nucl.\ Phys.\  {\bf B422}, 37 (1994).

\bibitem{Laenen:1993xs}
E.~Laenen, S.~Riemersma, J.~Smith and W.~L.~van Neerven,
Nucl.\ Phys.\  {\bf B392}, 229 (1993).

\bibitem{Berger:1979du}
E.~L.~Berger and S.~J.~Brodsky,
Phys.\ Rev.\ Lett.\  {\bf 42}, 940 (1979);\\
J.~F.~Gunion, P.~Nason and R.~Blankenbecler,
Phys.\ Rev.\  {\bf D29}, 2491 (1984).

\bibitem{Abbott:1980as}
L.~F.~Abbott and R.~M.~Barnett,
Annals Phys.\  {\bf 125}, 276 (1980);\\
L.~F.~Abbott, W.~B.~Atwood and R.~M.~Barnett,
Phys.\ Rev.\  {\bf D22}, 582 (1980).

\bibitem{Buras:1980yt}
A.~J.~Buras,
Rev.\ Mod.\ Phys.\  {\bf 52}, 199 (1980).

\bibitem{Bednyakov:1984gh}
V.~A.~Bednyakov, I.~S.~Zlatev, Y.~P.~Ivanov, P.~S.~Isaev and S.~G.~Kovalenko,
Sov.\ J.\ Nucl.\ Phys.\  {\bf 40}, 494 (1984).

\bibitem{Penin:1997zk}
A.~A.~Penin and A.~A.~Pivovarov,
Phys.\ Lett.\  {\bf B401}, 294 (1997)
[hep-ph/9612204].

\bibitem{Mahapatra:1997av}
B.~P.~Mahapatra, Preprint SU-PHY-97-03 (1997).

\bibitem{Adams:1999sx}
T.~Adams {\it et al.}  [NuTeV Collaboration],
Talk given at 34th Recontres de Moriond: QCD and Hadronic 
Interactions, Les Arcs, France, Mar 1999
[hep-ex/9906037].

\bibitem{Martin:1995kk}
A.~D.~Martin, W.~J.~Stirling and R.~G.~Roberts,
Phys.\ Rev.\  {\bf D51}, 4756 (1995)
[hep-ph/9409410].

\bibitem{Atwood:1973zp}
W.~B.~Atwood and G.~B.~West,
Phys.\ Rev.\  {\bf D7} (1973) 773.

\bibitem{Lacombe:1980dr}
M.~Lacombe, B.~Loiseau, J.~M.~Richard, R.~Vinh Mau, J.~Cote, P.~Pires 
and R.~De Tourreil,
Phys.\ Rev.\  {\bf C21}, 861 (1980);\\
M.~Lacombe, B.~Loiseau, R.~Vinh Mau, J.~Cote, P.~Pires and R.~de Tourreil,
Phys.\ Lett.\  {\bf B101}, 139 (1981).

\bibitem{Sokolov:1988mw}
S.~N.~Sokolov, Preprint IFVE-88-110 (1988).

\bibitem{Whitlow:1990dr}
L.~W.~Whitlow, Report SLAC-0357 (1990).

\bibitem{Giele:1998gw}
W.~T.~Giele and S.~Keller,
Phys.\ Rev.\  {\bf D58}, 094023 (1998)
[hep-ph/9803393].

\bibitem{Gross:1974fm}
D.~J.~Gross,
Phys.\ Rev.\ Lett.\  {\bf 32}, 1071 (1974).

\bibitem{Seligman:1997fe}
W.~G.~Seligman, Report NEVIS-292 (1997).

\bibitem{Kataev:2000dp}
A.~L.~Kataev, G.~Parente and A.~V.~Sidorov,
Nucl.\ Phys.\  {\bf A666}, 184 (2000)
[hep-ph/9907310].

\bibitem{Werlen:1999fn}
M.~Werlen, Preprint LAPTH-734-99 [hep-ph/9906483].

\bibitem{Laenen:2000ii}
E.~Laenen, G.~Sterman and W.~Vogelsang,
Contributed to 8th Inetrnational Workshop on Deep Inelastic Scattering
and QCD (DIS 2000), Liverpool, England, Apr 2000
[hep-ph/0006352].

\bibitem{Prytz:1993vr}
K.~Prytz,
Phys.\ Lett.\  {\bf B311}, 286 (1993).

\bibitem{Seligman:1997mc}
W.~G.~Seligman {\it et al.},
Phys.\ Rev.\ Lett.\  {\bf 79}, 1213 (1997).

\bibitem{Bodek:2000yr}
U.~K.~Yang {\it et al.}  [CCFR/NuTeV Collaboration],
hep-ex/0009041 (2000). 

\bibitem{Gomez:1994ri}
J.~Gomez {\it et al.},
Phys.\ Rev.\  {\bf D49} (1994) 4348.

\bibitem{Melnitchouk:1999un}
W.~Melnitchouk, I.~R.~Afnan, F.~Bissey and A.~W.~Thomas,
Phys.\ Rev.\ Lett.\  {\bf 84}, 5455 (2000)
[hep-ex/9912001].

\bibitem{Yang:2000ew}
U.~K.~Yang and A.~Bodek,
Phys.\ Rev.\ Lett.\  {\bf 84}, 5456 (2000)
[hep-ph/9912543].

\bibitem{Bethke:2000ai}
S.~Bethke,
J.\ Phys.\ G {\bf G26}, R27 (2000)
[hep-ex/0004021].

\bibitem{Arneodo:1993kz}
M.~Arneodo {\it et al.}  [New Muon Collaboration],
Phys.\ Lett.\  {\bf B309}, 222 (1993).

\bibitem{Alekhin:2000af}
S.~I.~Alekhin and A.~L.~Kataev,
Nucl.\ Phys.\  {\bf A666-667}, 179 (2000)
[hep-ph/9908349].

\bibitem{Szczurek:2000wp}
A.~Szczurek and V.~Uleshchenko,
Phys.\ Lett.\  {\bf B475}, 120 (2000)
[hep-ph/9911467];\\
A.~Szczurek,
Talk given at 8th International Workshop on Deep Inelastic Scattering
and QCD (DIS 2000), Liverpool, England, Apr 2000
[hep-ph/0006320].

\bibitem{QCD} 
A.~Ahmadov, {\it et al.},
Proceedings of the Workshop on Standard Model Physics
(and more) at the LHC, ed. G.Altarelli and M.L.Mangano, CERN-2000-04 (2000);//
S.~Catani {\it et al.}, hep-ph/0005025.


\end{thebibliography}
\end{document}